\documentclass[aps,pra,twocolumn,showpacs,floatfix,nofootinbib,groupedaddress]{revtex4}

\usepackage{amsmath,amssymb,amsfonts,graphics,graphicx,times,bm,bbm,color}
\usepackage{natbib}
\usepackage{hyperref}
\usepackage{hypernat}

\newcommand{\ignore}[1]{}

\begin{document}

\title{Accuracy \textit{vs} run time in adiabatic quantum search}
\author{A. T. Rezakhani$^{(1,4)}$, A. K. Pimachev$^{(1,4)}$, and D. A. Lidar$^{(1,2,3,4)}$}
\affiliation{$^{(1)}$Departments of Chemistry, $^{(2)} $Physics, and $^{(3)}$Electrical
Engineering, and $^{(4)}$Center for Quantum Information Science \&
Technology, University of Southern California, Los Angeles, California
90089, USA}

\begin{abstract}
Adiabatic quantum algorithms are characterized by their run time and
accuracy. The relation between the two is essential for quantifying
adiabatic algorithmic performance, yet is often poorly understood. We
study the dynamics of a continuous time, adiabatic quantum search
algorithm, and find rigorous results relating the accuracy and the run
time. Proceeding with estimates, we show that under fairly general circumstances the adiabatic
algorithmic error exhibits a behavior with two discernible regimes:
the error decreases exponentially for short times, then decreases
polynomially for longer times. We show that the well known quadratic speedup over
classical search is associated only with the exponential error regime. We illustrate the results through examples of evolution paths derived by minimization of the adiabatic error. We also discuss
specific strategies for controlling the adiabatic error and run time.
\end{abstract}


\maketitle


\section{Introduction}

\label{Int}

In adiabatic quantum computation (AQC) \cite{Brooke:99,Farhi1,Farhi2}
quantum algorithms are implemented by initializing a system in an easily
prepared ground state, followed by adiabatic evolution subject to a
Hamiltonian whose final ground state represents the solution to a
computational problem. It is known that this model is computationally
equivalent to the standard circuit model of quantum computation, in the
sense that each model can simulate the other with polynomial resource
overhead \cite{AharonovAdiabatic,Oliveira:05,MLM}. While error correction
methods have been proposed for AQC \cite{Jordan:05,Lidar:AQC-DD}, and
arguments have been put forth that AQC\ is inherently insensitive to certain
types of errors \cite{Childs:01,SarandyLidar:05,AminThermalAQC}, unlike the
circuit model it is still an open question whether AQC\ can be made fault
tolerant subject to realistic noise models pertaining to AQC\ in an open
system setting. Indeed, even what constitutes a consistent picture of adiabatic
evolution in open systems is still the subject of some debate
\cite{SarandyLidar:04,Oreshkov:10}. Nevertheless, experiments in AQC using
superconducting qubits have made great strides recently \cite{2010arXiv1004.1628H}.

The performance of adiabatic quantum algorithms is characterized by the
\textquotedblleft adiabatic error\textquotedblright , i.e., the fidelity
loss between the actual time-evolved state (the solution of the Schr\"{o}%
dinger equation) and the instantaneous ground state, calculated at the final
time $T$. Since the adiabatic theorem \cite%
{Fock,Kato,Messiah:book,Teufel:book} guarantees that in the limit of
arbitrarily slow evolution the error approaches zero, one expects that a
more slowly varying Hamiltonian and/or a longer evolution time should result
in higher fidelity or accuracy. This is the \emph{accuracy-time tradeoff} in
quantum adiabatic algorithms. This tradeoff is often formalized by the
adiabatic condition, which states (roughly) that the variation rate of the
Hamiltonian should be $\ll $ the adiabatic error times the gap squared. A
clear disadvantage of this condition is the inherent vagueness of
\textquotedblleft $\ll $,\textquotedblright\ which makes it difficult to
reliably quantify the evolution time \textit{vs} the desired accuracy.
Moreover, violations of the traditional adiabatic condition have been
reported \cite{MarzlinSanders,Tong:05,du:060403}, in the sense that the
condition is neither necessary nor sufficient for adiabatic evolution. While
these violations have been explained as being either due to inconsistent
manipulations \cite{SarandyWuLidar:04} or due to resonant transitions \cite%
{Amin}, it appears that there is no single \textquotedblleft adiabatic
theorem\textquotedblright . Rather, a number of different rigorous
conditions have been derived, which apply under various mathematical
assumptions
\cite{Avron:871,Avron:872,Nenciu:93,Martinez,HagedornJoye:02,JRS,o'hara:042319,Boixo}.
A rigorous condition which holds for analytic Hamiltonians and exibits
the explicit scaling with system size, needed for AQC resource quantification,
was derived in Ref.~\cite{LRH}. 

In this work we perform a comprehensive analysis of the adiabatic error for
the case of an adiabatic quantum search algorithm, in a closed system
setting. We focus on quantum search not only because it is an
important example of a quantum speedup \cite{Grover}, but also because
it is amenable to an exact analytical treatment. Indeed, rather than relying on a
particular form of the adiabatic theorem, our approach is based on an exact
treatment of the underlying dynamics. We calculate the adiabatic error as an
explicit function of the evolution time. We work out the formal $1/T$
expansion of the error---often used in rigorous treatments of the adiabatic
theorem---through which we provide a large-$T$ \emph{polynomial} upper bound
for the adiabatic error. We shall argue that a leading order truncation of
this series expansion can result in misleading estimates for the scaling of
the evolution time \textit{vs} system size. We draw this conclusion on the
basis of a careful study of the adiabatic error, showing that it 
may exhibit 
a short-$T$ \emph{exponential }decay, which is hidden in the formal polynomial
expansion. We show that this short-time exponential decay heralds the
correct scaling for the evolution time \textit{vs} system size (a quadratic
speedup over classical search), avoiding the overestimation that results
from the long-time polynomial decay, and which leads to a loss of the quantum
speedup. Additionally, we propose specific adiabatic evolution paths (``interpolations'')---inspired by
a procedure for minimization of time functionals for adiabatic
evolution \cite{QAB}---and use these to illustrate our results. We
also examine the applicability of the 
traditional measure of the evolution time and contrast this to what we
obtain from our analysis. This careful analysis enables us to show explicitly
how one can reduce the adiabatic error as a function of the evolution time.

The structure of this paper is as follows. We start with some general
background in Sec.~\ref{Ad-Int} by delineating the framework of adiabatic
quantum computation, the definition of the adiabatic error, and the
adiabatic theorem. In Sec.~\ref{QSHam}, we specialize to the specific
problem of quantum search. There we introduce a general Hamiltonian
interpolation for the search problem, and analytically solve the corresponding Schr\"{o}%
dinger equation. Section~\ref{AdError} contains the core of our results. We
first derive an exact expression for the adiabatic error in subsection~\ref%
{exactr}. Next, in subsection~\ref{Approx}, we approximate this expression
and show how the polynomial and exponential behaviors emerge. Since
estimation of the adiabatic error requires specific interpolation paths, we
derive a general interpolation for the search Hamiltonian in subsection~\ref%
{HAM}, and investigate in detail three special cases. The adiabatic error
for the general interpolation together with the special cases is estimated
in subsection~\ref{ErPaths}. In subsection~\ref{strategy}, 
we discuss a
strategy for controllably reducing the adiabatic error. There we show how one
can employ a freedom in the interpolation to manipulate the adiabatic error.
This accounts for the performance-resource tradeoff. We conclude with a
summary of the results and an outlook in Sec.~\ref{Summ}.


\section{Framework of adiabatic quantum computation}

\label{Ad-Int}

We begin by defining the adiabatic error as it arises in the context of AQC.
We also provide a brief review of some pertinent facts concerning the
\textquotedblleft traditional\textquotedblright\ adiabatic theorem, also in
the context of AQC. However, we note that in the sequel we shall not use the
adiabatic theorem; rather, we shall treat the dynamics directly by solving
the Schr\"{o}dinger equation, and later enforce adiabaticity by means of a $%
1/T$ expansion.


\subsection{Adiabatic error}

Let us treat the total evolution time $T$ as a parameter and define the the
scaled (dimensionless) time%
\begin{equation}
\tau =t/T\in \lbrack 0,1].
\end{equation}%
Assume that we have an $N$-dimensional quantum system that evolves for a
total time $T$ under the Hamiltonian $H_{T}(\tau )$ with spectral
decomposition 
\begin{equation}
H_{T}(\tau )=\sum_{i=0}^{N-1}E_{i}(\tau )|\Phi _{i}(\tau )\rangle \langle
\Phi _{i}(\tau )|,~~\tau \in \lbrack 0,1]  \label{H}
\end{equation}%
where $E_{i}(\tau )\leq E_{j}(\tau )$ for $i<j$, except that the ground
state energy is separated from the rest of the spectrum by a nonvanishing
gap 
\begin{equation}
D(\tau )\equiv E_{1}(\tau )-E_{0}(\tau )>0.
\end{equation}%
Possible level crossing among excited eigenstates $\{|\Phi _{i}(\tau
)\rangle \}_{i>0}$ shall not concern us here because in the following we
shall only focus on the ground state $|\Phi _{0}(\tau )\rangle $. We assume
that the system is closed, i.e., the evolution is governed by the Schr\"{o}%
dinger equation 
\begin{equation}
i|\dot{\psi}_{T}(\tau )\rangle =TH_{T}(\tau )|\psi _{T}(\tau )\rangle ,
\label{eqscalt}
\end{equation}%
where from now on dot denotes $\mathrm{d}/\mathrm{d}\tau $, and we set $%
\hbar \equiv 1$. The initial state is assumed to be the ground state, i.e., 
\begin{equation}
|\psi _{T}(0)\rangle =|\Phi _{0}(0)\rangle .
\end{equation}

\textit{Remark}. We shall use the subscript $T$ to indicate the parametric
dependence on $T$. When symbols already have a subscript we shall avoid the
additional $T$ subscript so as not to clutter the notation.

A central quantity of interest to us is the \textquotedblleft adiabatic
error\textquotedblright\ $\delta _{\text{ad}}(\tau )$, which quantifies the
distance between the instantaneous ground state and the actual state: 
\begin{equation}
\delta _{\text{ad}}(\tau )\equiv \sqrt{1-|\langle \Phi _{0}(\tau )|\psi
_{T}(\tau )\rangle |^{2}}.  \label{def:ad-err}
\end{equation}%
Note that $\delta _{\text{ad}}(\tau )$ is a distance and (in)fidelity
measure in their rigorous sense. Indeed, the trace distance $\mathcal{D}$
and fidelity $\mathcal{F}$ between two arbitrary density operators $\varrho
_{1}$ and $\varrho _{2}$ are defined as \cite{Nielsen:book} 
\begin{eqnarray}
\mathcal{D}(\varrho _{1},\varrho _{2}) &\equiv &\frac{1}{2}\Vert \varrho
_{1}-\varrho _{2}\Vert _{1},  \label{tr-dist} \\
\mathcal{F}(\varrho _{1},\varrho _{2}) &\equiv &\Vert \sqrt{\varrho _{1}}%
\sqrt{\varrho _{2}}\Vert _{1},  \label{fid}
\end{eqnarray}%
where $\Vert X\Vert _{1}\equiv \mathrm{Tr}\sqrt{X^{\dag }X}$ is the trace
norm (sum of singular values of $X$). For pure states $\varrho _{i}=|\psi
_{i}\rangle \langle \psi _{i}|$ it is not hard to show that $\mathcal{D}%
(\varrho _{1},\varrho _{2})=\sqrt{1-|\langle \psi _{1}|\psi _{2}\rangle |^{2}%
}\equiv \mathcal{D}(|\psi _{1}\rangle ,|\psi _{2}\rangle )$ and $\mathcal{F}%
(\varrho _{1},\varrho _{2})=|\langle \psi _{1}|\psi _{2}\rangle |\equiv 
\mathcal{F}(|\psi _{1}\rangle ,|\psi _{2}\rangle )$. Thus%
\begin{eqnarray}
\delta _{\text{ad}}(\tau ) &=&\mathcal{D}(|\psi _{T}(\tau )\rangle ,|\Phi
_{0}(\tau )\rangle )  \notag \\
&=&\sqrt{1-\mathcal{F}(|\psi _{T}(\tau )\rangle ,|\Phi _{0}(\tau )\rangle
)^{2}}.
\label{delta-D}
\end{eqnarray}

An equivalent and useful formulation of $\delta _{\text{ad}}(\tau )$ is the
following. Let $P(\tau )\equiv |\Phi _{0}(\tau )\rangle \langle \Phi
_{0}(\tau )|$ denote the ground-state eigenprojection. The connection
between the initial preparation $P(0)$ and the time evolved state $P(\tau )$
is given by 
\begin{equation}
P(\tau )=A(\tau )P(0)A^{\dag }(\tau ),  \label{ad-int}
\end{equation}%
where the unitary operator $A(\tau )$---called the adiabatic intertwiner 
\cite{Avron:871}---determines the evolution in the eigenspace $P(\tau )$ and
its orthogonal complement $\openone-P(\tau )$: 
\begin{equation}
A(\tau )\equiv \sum_{i=0}^{N-1}|\Phi _{i}(\tau )\rangle \langle \Phi
_{i}(0)|.
\end{equation}%
One can assign a (dimensionless) \textquotedblleft adiabatic
Hamiltonian,\textquotedblright\ $H_{A}(\tau )$, to this evolution defined
via 
\begin{equation}
i\dot{A}(\tau )=H_{A}(\tau )A(\tau ).  \label{ad-schr}
\end{equation}%
On the other hand, the evolution operator generated as 
\begin{equation}
i\dot{V}_{T}(\tau )=TH(\tau )V_{T}(\tau )
\end{equation}%
dictates the actual dynamics: 
\begin{equation}
|\psi _{T}(\tau )\rangle \langle \psi _{T}(\tau )|=V_{T}(\tau
)P(0)V_{T}^{\dag }(\tau ).
\end{equation}%
Thus the error $\delta _{\text{ad}}(\tau )$ in fact measures the difference
between $V_{T}(\tau )$ and $A(\tau )$, or equivalently how far the unitary
operator 
\begin{equation}
\Omega _{T}(\tau )\equiv A^{\dag }(\tau )V_{T}(\tau )  \label{OMEGA}
\end{equation}%
is from $\openone$ (identity). Indeed, we have 
\begin{equation}
\delta _{\text{ad}}(\tau )=\sqrt{1-|\langle \Phi _{0}(0)|\Omega _{T}(\tau
)|\Phi _{0}(0)\rangle |^{2}},  \label{def:ad-err2}
\end{equation}%
which vanishes iff $\Omega _{T}(\tau )=\openone$.

The above formulations have presumed that the degeneracy of the ground-state
eigensubspace does not change in time. Nonetheless, there may be situations
in which this degeneracy preservation assumption does not hold. In fact, as
we shall see later, the quantum search problem we study in this paper falls
into this category. Let us assume that the initial state is pure, $|\psi
_{T}(0)\rangle =|\Phi _{0}(0)\rangle $, but the instantaneous ground-state
eigenprojection $P(\tau )$ is not necessarily rank-$1$, accounting for the
possibility of degeneracy. Intuitively, if the actual state $|\psi _{T}(\tau
)\rangle $ only has components in the support of $P(\tau )$ the algorithm
has achieved its goal at the instant $\tau $, whereas a less than full
overlap denotes lack of success at this instant. The overlap of $|\psi
_{T}(\tau )\rangle $ and the support of $P(\tau )$ can be quantified, e.g.,
with $\langle \psi _{T}(\tau )|P(\tau )|\psi _{T}(\tau )\rangle $, whence we
define the following performance error: 
\begin{equation}
\delta _{\text{ad}}^{\prime }(\tau )\equiv \sqrt{1-\langle \psi _{T}(\tau
)|P(\tau )|\psi _{T}(\tau )\rangle }.  \label{error-final-form}
\end{equation}%
It is evident that $0\leq \delta _{\text{ad}}^{\prime }(\tau )\leq 1$, with $%
\delta _{\text{ad}}^{\prime }=1$ iff the actual state has vanishing overlap
with the support of the instantaneous ground-state eigenprojection, while $%
\delta _{\text{ad}}^{\prime }=0$ iff the actual state resides anywhere in
the support.

We can write $\langle \Phi _{0}(0)|\Omega _{T}(\tau )|\Phi _{0}(0)\rangle =%
\mathrm{Tr}[P(0)\Omega _{T}(\tau )]$ and $\langle \psi _{T}(\tau )|P(\tau
)|\psi _{T}(\tau )\rangle =\mathrm{Tr}[P(\tau )V_{T}(\tau )P(0)V_{T}^{\dag
}(\tau )]=\mathrm{Tr}[P(0)\Omega _{T}(\tau )P(0)\Omega _{T}^{\dag }(\tau )]$%
. Using this, note the difference between 
\begin{equation}
\delta _{\text{ad}}(\tau )=\sqrt{1-|\mathrm{Tr}[P(0)\Omega _{T}(\tau )]|^{2}}
\end{equation}%
and 
\begin{equation}
\delta _{\text{ad}}^{\prime }(\tau )=\sqrt{1-\mathrm{Tr}[P(0)\Omega
_{T}(\tau )P(0)\Omega _{T}^{\dag }(\tau )]}.
\end{equation}%
Namely, in the nondegenerate case $\delta _{\text{ad}}(\tau )=0$ iff $\Omega
_{T}(\tau )=\openone$, indicating that the evolution was perfectly
adiabatic. In the degenerate case, on the other hand, algorithmic success
does not require the evolution to be perfectly adiabatic since only
nonvanishing overlap with the ground-state eigensubspace is required, i.e., $%
\delta _{\text{ad}}^{\prime }=0$ iff $V_{T}(\tau )P(0)V_{T}^{\dag
}(\tau )\in \mathrm{supp}[P(\tau )]$, whereas complete failure
requires 
the dynamics to remove any overlap with the ground-state eigensubspace,
i.e., $\delta _{\text{ad}}^{\prime }=1$ iff $V_{T}(\tau )P(0)V_{T}^{\dag }(\tau )
\notin \mathrm{supp}[P(\tau )]$.

While the error as defined in Eq. (\ref{error-final-form}) is not
necessarily a distance in the strict sense, it is adequate for quantifying
the adiabatic error. Note also that $\delta _{\text{ad}}^{\prime }(\tau )$
reduces to $\delta _{\text{ad}}(\tau )$ when the ground state is
nondegenerate. From now on we shall be using these various expressions for
the adiabatic error as appropriate in the rest of this paper.


\subsection{The adiabatic theorem}

One variant of the \textquotedblleft traditional\textquotedblright\
adiabatic theorem \cite{Messiah:book} states that given an $0<\varepsilon
\ll 1$ and a time-dependent Hamiltonian $H(\tau )$ with a nondegenerate
ground state, the adiabatic error satisfies $\delta _{\text{ad}}(1)\leq
\varepsilon $, provided that 
\begin{equation}
\frac{\max_{\tau }\Vert \dot{H}(\tau )\Vert }{\min_{\tau }D^{2}(\tau )}\ll
\varepsilon T,  \label{cond}
\end{equation}%
in which $\Vert \cdot \Vert $ is the standard operator norm, defined as the
maximum singular value, i.e.,
\begin{equation}
  \Vert X\Vert \equiv \sup_{|v\rangle ,\Vert
    v\Vert =1}|\langle v|\sqrt{X^{\dag }X}|v\rangle |,
\label{op-norm}
\end{equation}
which reduces to $%
\sup_{|v\rangle ,\Vert v\Vert =1}|\langle v|X|v\rangle |$ for normal
operators. As remarked in the Introduction, 
this condition is hardly
quantitative due to the intrinsic vagueness of \textquotedblleft $\ll $%
\textquotedblright , and has been the subject of critique (consistent with
its lack of rigor) \cite{MarzlinSanders,Tong:05,du:060403}, justifications 
\cite{SarandyWuLidar:04,Amin}, and rigorous improvements \cite%
{Avron:871,Avron:872,Nenciu:93,Martinez,HagedornJoye:02,JRS,o'hara:042319,Boixo,LRH}%
. Nevertheless, it remains a useful rule of thumb, as long as it is applied
with appropriate care.

An immediate implication of the adiabatic theorem is that, assuming it
is initialized in the
ground state, the system remains close to the final ground state at $t=T$.
Thus, by choosing the Hamiltonian such that $H(0)$ corresponds to a \textit{%
simple} ground state $|\Phi _{0}(0)\rangle $ (simple in the sense that it is
easily preparable), and $H(1)$ represents a Hamiltonian whose ground state $%
|\Phi _{0}(1)\rangle $ identifies the solution to a computationally \textit{%
hard} problem, one can devise an adiabatic version for the corresponding
algorithmic or computational task. This is precisely the insight that led to
the advent of AQC \cite{Farhi1,Farhi2}.

A simple ``annealing schedule'', or ``path'', between $H(0)$ and $H(1)$ is the following linear
interpolation in $\tau $: 
\begin{equation}
H(\tau )=(1-\tau )H(0)+\tau H(1).  \label{aaa}
\end{equation}%
In physical situations, however, one often realizes the dynamics by tuning
some time-dependent control knobs or couplings $\mathbf{x}(\tau )=\bigl(%
x_{1}(\tau ),\ldots ,x_{K}(\tau )\bigr)$ of the Hamiltonian. This suggests
that a generalization of Eq. (\ref{aaa}) can be introduced by assuming
access to a controllable set of non-commuting, linearly-independent
primitive Hamiltonians $\{H_{i}\}_{i=1}^{K\leq N}$ combined, e.g., as $H[%
\mathbf{x}(\tau )]=\sum_{i=1}^{K}x_{i}(\tau )H_{i}$. Further generalizations
can be introduced as well \cite{Farhi1,RolandCerf,QAB,Andrecut}. However,
for our purposes in this paper we shall consider the parametrization \cite%
{RolandCerf,QAB} 
\begin{equation}
H\bigl(\mathbf{x}(\tau )\bigr)=x_{1}(\tau )H(0)+x_{2}(\tau )H(1),
\label{2-d-interpol}
\end{equation}%
with the boundary conditions 
\begin{eqnarray}
&&\bigl(x_{1}(0),x_{2}(0)\bigr)=(1,0),  \label{BC-1} \\
&&\bigl(x_{1}(1),x_{2}(1)\bigr)=(0,1).  \label{BC-2}
\end{eqnarray}

Two remarks are in order regarding AQC. (i) A primary goal in 
AQC is to make $\delta _{\text{ad}}(1)$ decrease more
rapidly for a given $T$ and problem size, or alternatively, to make $T$
smaller for a given $\varepsilon $ and problem size. Often the problem size
is given by $N$, the dimension of the Hilbert space. However, in the context
of many-body quantum systems, where the Hilbert space is a tensor product of
subsystems (e.g., qubits), $\log N$ is the correct measure of problem size,
coinciding with system size. (ii) The \textquotedblleft run
time\textquotedblright\ complexity of a quantum algorithm should be defined
as 
\begin{equation}
\tau _{\text{run}}\equiv T\times \max_{\tau }\Vert H(\tau )\Vert ,
\label{runT}
\end{equation}%
not $T$ \cite{AharonovAdiabatic}. This \textit{regularization} is required
because of the energy-time tradeoff in quantum mechanics, in the sense that
multiplication of $H(\tau )$ in Eq.~(\ref{cond}) by some positive factor $%
\alpha $ manifests 
itself as dividing $T$ by the same factor, making it
possible to decrease $T$ arbitrarily by choosing $\alpha $ sufficiently
large. This tradeoff can also be understood via the Schr\"{o}dinger equation
(\ref{eqscalt}), in which the final state of a system evolving under
Hamiltonian $H(\tau )$ for $T$ is the same as that of a system evolving
under $\alpha H(\tau )$ for $T/\alpha $. This ambiguity is fixed by the
definition of $\tau _{\text{run}}$ as in Eq.~(\ref{runT}). Scaling of $\tau
_{\text{run}}$ with system size, for a given upper bound on the error 
$\delta _{\text{ad}}(1) \leq \varepsilon$, 
determines the run time complexity of the corresponding quantum algorithm.


\section{Quantum search Hamiltonian}

\label{QSHam}

Grover's quantum search algorithm \cite{Grover} performs a search for $M$
\textquotedblleft marked\textquotedblright\ items among $N$ items of an
unsorted database, presuming that there is an \textquotedblleft
oracle\textquotedblright\ for distinguishing the marked from the unmarked
items. The algorithm in its original form ($M=1$) comprises the following
steps: (i) assign orthonormal quantum states (i.e., labels) $\{|0\rangle
,\ldots ,|N-1\rangle \}$ to the items, (ii) prepare the quantum system in
the equal superposition state $\sum_{i=0}^{N-1}|i\rangle /\sqrt{N}$, and
(iii) apply the \textquotedblleft Grover operator\textquotedblright
---encompassing the oracle---repeatedly \cite{Grover,Nielsen:book}. The
algorithm finds a marked item after $\tau _{\text{run}}=O(\sqrt{N/M})$ calls
of the oracle---a quadratic speedup over the best classical algorithm---and
is provably optimal 
for any $N$ (not necessarily very large) \cite{Zalka:97}. Various
generalization of the algorithm 
have been introduced (e.g.,
Refs.~\cite{Grover-2,Boyer:96,Grover-3,G-1,Lidar:PRA01Grover,G-2}),
and it has also been implemented experimentally in a number of
physical settings (e.g., Refs.~\cite{Chuang,Jones,Ollerenshaw:02,Daems,Ivanov}).

An adiabatic Hamiltonian version of the search algorithm was first
introduced in Ref.~\cite{Farhi1}, 
but failed to display the expected quadratic
speedup as it relied on the linear interpolation of Eq. (\ref{aaa}). This
was fixed in Ref.~\cite{RolandCerf} 
by using a non-linear but one-dimensional
interpolation with $x_{2}(\tau )=1-x_{1}(\tau )$ [recall Eq.~(\ref%
{2-d-interpol})], which moves fast when away from the minimum gap, but slows
down near it. As shown in Ref.~\cite{QAB}, 
this result can be further improved,
in the sense of a smaller adiabatic error, by adopting a two-dimensional
interpolation as in Eq.~(\ref{2-d-interpol}), 
\begin{equation}
H(\tau )/J=x_{1}(\tau )H_{\mathcal{I}}+x_{2}(\tau )H_{\mathcal{M}},
\label{search-Ham}
\end{equation}%
with the two projective Hamiltonians 
\begin{eqnarray}
&&H_{\mathcal{I}}=\openone-|\phi \rangle \langle \phi |, \\
&&H_{\mathcal{M}}=\openone-P_{\mathcal{M}},
\end{eqnarray}%
where $|\phi \rangle \equiv \sum_{i=0}^{N-1}|i\rangle /\sqrt{N}$ is the
equal superposition of all of the \textquotedblleft label\textquotedblright\
states (items), $P_{\mathcal{M}}\equiv \sum_{m\in \mathcal{M}}|m\rangle
\langle m|$ is the projection over the subspace $\mathcal{M}$ of the marked
items ($|\mathcal{M}|=M$), and $J$ is a dimensional constant which sets the 
energy scale. In other words, the initial state $|\phi \rangle $ is the ground
state of the initial Hamiltonian $H_{\mathcal{I}}$, while any state
supported fully on $\mathcal{M}$ is a ground state of the final (oracle)
Hamiltonian $H_{\mathcal{M}}$. Note that unlike previous treatments of
adiabatic quantum search \cite{Farhi1,RolandCerf,QAB}, the Hamiltonian $H_{%
\mathcal{M}}$ has a degenerate ground eigenspace spanned by $\{|m\rangle
\}_{m\in \mathcal{M}}$.

We remark that the search Hamiltonian (\ref{search-Ham}) is a member of the
following class of projective Hamiltonians \cite%
{QAB,AQC-intrinsic,AharonovTa-Shma}: 
\begin{equation}
H\bigl(\mathbf{x}(\tau )\bigl)=x_{1}(\tau )P_{\mathbf{a}}^{\perp
}+x_{2}(\tau )P_{\{\mathbf{b}\}}^{\perp },  \label{general-Ham}
\end{equation}%
where $P_{\mathbf{a}}^{\perp }\equiv \openone-|\mathbf{a}\rangle \langle 
\mathbf{a}|$, $P_{\{\mathbf{b}\}}^{\perp }\equiv \openone-\sum_{\mathbf{b}}|%
\mathbf{b}\rangle \langle \mathbf{b}|$, with $|\mathbf{a}\rangle $ and $\{|%
\mathbf{b}\rangle \}$ fixed (normalized) vectors in the system Hilbert
space, for which $\langle \mathbf{a}|P_{\{\mathbf{b}\}}^{\perp }|\mathbf{a}%
\rangle $ is a given function of $N$. In the case of the search problem, we
have $|\mathbf{a}\rangle =|\phi \rangle $, $|\mathbf{b}\rangle =|m\rangle $,
whence $|\langle \mathbf{a}|\mathbf{b}\rangle |=1/\sqrt{N}$. The results of
this paper can be generalized to other members of the class of projective
Hamiltonians.


\subsection{Two-dimensional reduction}

In the computational basis, in which $|\psi _{T}(\tau )\rangle
=\sum_{i=0}^{N-1}\psi _{i}(\tau )|i\rangle $, the Schr\"{o}dinger equation (%
\ref{eqscalt}) becomes 
\begin{equation}
i\dot{\psi _{i}}=T\left[ \left( x_{1}+x_{2}-x_{2}\sum_{m\in \mathcal{M}%
}\delta _{mi}\right) \psi _{i}-\frac{x_{1}}{N}\sum_{j=1}^{N}\psi _{j}\right]
,
\end{equation}%
with the initial value $\psi _{i}(0)=1/\sqrt{N}$. It can be seen from
this expression that the marked
components all behave similarly, as do the unmarked components. Hence we can
rewrite the state $|\psi _{T}(\tau )\rangle $ as 
\begin{equation}
|\psi _{T}(\tau )\rangle =\psi _{\text{u}}(\tau )\sum_{i\notin \mathcal{M}%
}|i\rangle +\psi _{\text{m}}(\tau )\sum_{i\in \mathcal{M}}|i\rangle ,
\end{equation}%
where the subscripts \textquotedblleft u\textquotedblright\ and
\textquotedblleft m\textquotedblright\ denote \textquotedblleft
unmarked\textquotedblright\ and \textquotedblleft marked,\textquotedblright\
respectively. The normalization condition now reads 
\begin{equation}
(N-M)|\psi _{\text{u}}(\tau )|^{2}+M|\psi _{\text{m}}(\tau )|^{2}=1.
\label{normalization}
\end{equation}%
%
%
%
%
By defining the (unnormalized) two-dimensional vector 
\begin{equation}
|\widetilde{\psi }\rangle =(\psi _{\text{u}},\psi _{\text{m}})^{T},
\end{equation}%
and the (non-Hermitian) reduced Hamiltonian matrix 
\begin{equation}
\widetilde{H}/J=\left( 
\begin{array}{cc}
rx_{1}+x_{2} & -rx_{1} \\ 
(r-1)x_{1} & (1-r)x_{1}%
\end{array}%
\right) ,
\label{pseudoHam}
\end{equation}%
in which 
\begin{equation}
r=M/N
\end{equation}%
is the fraction of the marked items, the Schr\"{o}dinger equation reduces to 
\begin{equation}
i|\dot{\widetilde{\psi }}_{T}(\tau )\rangle =T\widetilde{H}(\tau )|%
\widetilde{\psi }_{T}(\tau )\rangle ,
\label{sch-1}
\end{equation}%
with the initial condition $|\widetilde{\psi }(0)\rangle =(1/\sqrt{N},1/%
\sqrt{N})$. Therefore, not only is the parameter space of the problem
two-dimensional, it is described by an effectively two-dimensional
Hamiltonian (in the m-u representation). This reduction from the real
Hamiltonian $H$ [Eq.~(\ref{search-Ham})] to the effective Hamiltonian $%
\widetilde{H}$ [Eq.~(\ref{pseudoHam})] will prove useful in our analysis
below.

Later in the paper we shall need the norm of the Hamiltonian as well. This
can be calculated easily from Eqs.~(\ref{op-norm}) and
(\ref{pseudoHam}), and yields:
\begin{equation}
\Vert H(\tau )\Vert =%
\begin{cases}
J|x_1(\tau)+x_2(\tau)|:~~x_1(\tau)+x_2(\tau)\neq0, \\ 
J\sqrt{1-r} |x_1(\tau)|~~:~~\text{otherwise},%
\end{cases}
\label{Hnorm}
\end{equation}%
and similarly, 
\begin{equation}
\Vert \dot{H}(\tau )\Vert =%
\begin{cases}
J|\dot{x}_1(\tau)+\dot{x}_2(\tau)|:~~\dot{x}_1(\tau)+\dot{x}_2(\tau)\neq0,
\\ 
J\sqrt{1-r} |\dot{x}_1(\tau)|~~:~~\text{otherwise}.%
\end{cases}
\label{Hdotnorm}
\end{equation}


\subsection{Diagonalization and unitary interpolation}

The Hamiltonian $H/J$ (\ref{search-Ham}) has three distinct dimensionless
eigenvalues $E_{-}\leq E_{+}\leq E_{>}$, where 
\begin{eqnarray}
&&E_{\mp }=(x_{1}+x_{2}\mp \Delta )/2,  \label{E+_} \\
&&E_{>}=x_{1}+x_{2},  \label{E>}
\end{eqnarray}%
where 
\begin{equation}
\Delta \equiv \sqrt{(x_{1}-x_{2})^{2}+4rx_{1}x_{2}},  \label{gap}
\end{equation}%
is the dimensionless gap (hence $D\equiv J\Delta $) and $E_{>}$ is $(N-2)$%
-fold degenerate.

Let $\sigma _{z}=\mathrm{diag}(1,-1)$ and $\sigma _{y}=\left( 
\begin{smallmatrix}
0 & -i \\ 
i & 0%
\end{smallmatrix}%
\right) $ denote the Pauli matrices. Let us define the similarity matrix%
\begin{equation}
S\equiv \left( 
\begin{array}{cc}
r/\sqrt{1-r} & \sqrt{r} \\ 
-\sqrt{1-r} & \sqrt{r}%
\end{array}%
\right) ,  \label{formofS}
\end{equation}%
and the unitary%
\begin{equation}
\widehat{A}\equiv e^{-i\sigma _{y}\arccos [(x_{1}-(1-2r)x_{2})/\Delta ]/2}.
\label{formofA}
\end{equation}%
Then the effective Hamiltonian $\widetilde{H}(\tau )$ [Eq.~(\ref{pseudoHam})]
satisfies 
\begin{equation}
\widetilde{H}/J=S\widehat{H}(\tau )S^{-1},
\end{equation}%
where 
\begin{equation}
\widehat{H}(\tau )\equiv \widehat{A}~\mathrm{diag}(E_{+},E_{-})\widehat{A}%
^{\dag }  \label{Hhat}
\end{equation}%
is the Hermitian core of $\widetilde{H}/J$, and we easily find that
\begin{equation}
\widehat{H}(\tau )=\frac{1}{2}\left\{ [x_{1}(\tau )+x_{2}(\tau )]\openone%
+\Delta (\tau )\widehat{A}(\tau )\sigma _{z}\widehat{A}^{\dag }(\tau )\right\}
.  \label{uni-interp}
\end{equation}%
This last result is remarkable:\ it states that, up to an overall
(time-dependent) shift $[x_{1}(\tau )+x_{2}(\tau )]\openone$ and a conformal
factor $\Delta (\tau )$, the reduced Hamiltonian $\widehat{H}(\tau )$ is a 
\emph{unitary interpolation} $\widehat{A}(\tau )\sigma _{z}\widehat{A}^{\dag
}(\tau )$ \cite{Siu}. 
We will exploit this observation below.

The non-Hermitian reduced Hamiltonian $\widetilde{H}(\tau )$ and its
Hermitian core $\widehat{H}(\tau )$ have the same set of eigenvalues $%
\widehat{E}_{\mp }\equiv E_{\mp }$, and we have the spectral resolution%
\begin{equation}
\widehat{H}/J=\widehat{E}_{-}|\widehat{\Phi }_{-}\rangle \langle \widehat{%
\Phi }_{-}|+\widehat{E}_{+}|\widehat{\Phi }_{+}\rangle \langle \widehat{\Phi 
}_{+}|,  \label{wt-H}
\end{equation}%
where 
\begin{equation}
|\widehat{\Phi }_{\mp }(\tau )\rangle \equiv \widehat{A}(\tau )|z,\mp
\rangle ,  \label{phi-hat}
\end{equation}%
and $|\widehat{\Phi }_{\mp }(0)\rangle =|z,\mp \rangle $ are the
eigenvectors of $\sigma _{z}$, corresponding to the eigenvalues $\mp 1$. The
unitary operator $\widehat{A}(\tau )$ acts as a reduced adiabatic
intertwiner [Eq.~(\ref{ad-int})], in the sense that for the reduced
projection $\widehat{P}_{\mp }(\tau )\equiv |\widehat{\Phi }_{\mp }(\tau
)\rangle \langle \widehat{\Phi }_{\mp }(\tau )|$ we have 
\begin{equation}
\widehat{P}_{\mp }(\tau )=\widehat{A}(\tau )\widehat{P}_{\mp }(0)\widehat{A}%
^{\dag }(\tau ).
\end{equation}

\textit{Remark.} We emphasize that throughout the paper hat and tilde denote
states or operators in the reduced representation; the only exception is $%
\Delta $.


\subsection{Solving the Schr\"{o}dinger equation}

In solving the Schr\"{o}dinger equation and calculating the adiabatic error $%
\delta _{\text{ad}}(1)$, it is more convenient to work with the normalized
state 
\begin{equation}
|\widehat{\psi }_{T}(\tau )\rangle \equiv \sqrt{M}S^{-1}|\widetilde{\psi }%
_{T}(\tau )\rangle .  \label{psi-hat}
\end{equation}%
Equation~(\ref{sch-1}) now becomes 
\begin{equation}
i|\dot{\widehat{\psi }}_{T}(\tau )\rangle =T\widehat{H}(\tau )|\widehat{\psi 
}_{T}(\tau )\rangle ,
\end{equation}%
with $|\widehat{\psi }_{T}(0)\rangle =|z,-\rangle $. Solving this equation
results in 
\begin{equation}
|\widehat{\psi }_{T}(\tau )\rangle =\widehat{V}_{T}(\tau )|\widehat{\psi }%
_{T}(0)\rangle ,  \label{state-red}
\end{equation}%
in which 
\begin{equation}
\widehat{V}_{T}(\tau )\equiv \mathrm{Texp}\Bigl[-iT\int_{0}^{\tau }\widehat{H%
}(\tau ^{\prime })~\mathrm{d}\tau ^{\prime }\Bigr]
\end{equation}
is the time-ordered reduced evolution operator.


\subsubsection{General setup: Adiabatic interaction picture and Dyson series}

Having observed that the Grover search problem can be cast as a conformal
unitary interpolation, we outline a general, systematic approach for solving
the corresponding class of Schr\"{o}dinger equations using the Dyson series
expansion \cite{Sancho}. Consider as a specialization of the general
time-dependent Hamiltonian of Eq. (\ref{H}) the \textquotedblleft conformal
unitary interpolation Hamiltonian\textquotedblright\ 
\begin{equation}
H_{T}(\tau )=\Delta (\tau )A(\tau )H_{T}(0)A^{\dag }(\tau ),  \label{H_T}
\end{equation}%
in which 
\begin{equation}
H_{T}(0)=\sum_{i=0}^{N-1}E_{i}(0)P_{i}(0),
\end{equation}%
is the spectral decomposition of the (traceless) initial Hamiltonian $H(0)$,
the unitary operator $A(\tau )$ satisfies the adiabatic Schr\"{o}dinger
equation (\ref{ad-schr}) generated by the adiabatic Hamiltonian $H_{A}(\tau
) $ \cite{Avron:871,AQC-intrinsic}, and (the dimensionless gap) $\Delta
(\tau )>0$ is a smooth function with the initial value $\Delta (0)=1$. It
is also useful to think of the time-dependent Hamiltonian $H_{T}(\tau )$ [Eq. (%
\ref{H_T})] as the \textquotedblleft adiabatic interaction picture
Hamiltonian\textquotedblright , though normally an interaction picture
transformation does not involve a time-dependent prefactor such as $\Delta
(\tau )$. It is evident that the eigenvalues and eigenprojections of $%
H_{T}(\tau )$ satisfy 
\begin{eqnarray}
&&E_{i}(\tau )=\Delta (\tau )E_{i}(0),  \label{spect} \\
&&P_{i}(\tau )=A(\tau )P_{i}(0)A^{\dag }(\tau ).  \label{ap:P}
\end{eqnarray}%
Equation (\ref{spect}) implies that the spectrum of $H_{T}(\tau )$ evolves
conformally with $\Delta (\tau )$, i.e., all the eigenvalues are multiplied
by the same factor, while Eq.~(\ref{ap:P}) implies that the eigenprojections
are unitarily connected, and the degeneracy is constant in time. Comparison
of Eqs. (\ref{ad-int}) and (\ref{ap:P}) reveals that $P(\tau )=P_{0}(\tau )$
and here too, $A(\tau )$ plays the role of the adiabatic intertwiner.

Rather than solving the Schr\"{o}dinger equation 
\begin{equation}
\dot{V}_{T}(\tau )=TH_{T}(\tau )V_{T}(\tau ),  \label{Wdot}
\end{equation}%
we solve the equation of motion for the \textquotedblleft adiabatic
interaction picture\ propagator\textquotedblright\ 
[cf. Eq.~(\ref{OMEGA})]
\begin{equation}
\Omega _{T}(\tau )\equiv A^{\dag }(\tau )V_{T}(\tau ).
\end{equation}%
This provides a more direct tool for the calculation of the adiabatic error
[Eq.~(\ref{def:ad-err})]. Let us define 
\begin{eqnarray}
&&H_{0}(\tau )\equiv A^{\dag }(\tau )H_{T}(\tau )A(\tau )=\Delta (\tau
)H_{T}(0),  \label{H_0} \\
&&H_{1}(\tau )\equiv A^{\dag }(\tau )H_{A}(\tau )A(\tau ).  \label{H_1}
\end{eqnarray}%
Note that $H_{0}(\tau )$ has dimensions of energy while $H_{1}(\tau )$ is
dimensionless. In the adiabatic interaction picture the (dimensionless)
\textquotedblleft perturbation\textquotedblright\ is $TH_{0}(\tau
)-H_{1}(\tau )$, i.e., it follows from Eqs. (\ref{ad-schr}) and (\ref{Wdot})
that 
\begin{equation}
i\dot{\Omega}_{T}(\tau )=[TH_{0}(\tau )-H_{1}(\tau )]\Omega _{T}(\tau ).
\label{Omega-dott}
\end{equation}%
We also define the two unitaries $V_{0}(\tau )$ and $V_{1}(\tau )$ through
the following equations: 
\begin{eqnarray}
&&i\dot{V}_{0}=TH_{0}V_{0},  \label{V_0} \\
&&i\dot{V}_{1}=-V_{0}^{\dag }H_{1}V_{0}V_{1}.  \label{V_1}
\end{eqnarray}%
It is easily seen that $V_{0}V_{1}$ also satisfies Eq. (\ref{Omega-dott}),
so that 
\begin{equation}
\Omega _{T}(\tau )=V_{0}(\tau )V_{1}(\tau ).
\end{equation}

To simplify the analysis, we only consider Hamiltonians for which 
\begin{equation}
H_{A}(\tau )=h_{A}(\tau )~\Xi ,  \label{H_A}
\end{equation}%
in which $h_{A}$ is an integrable function and $\Xi $ is a constant ($\tau $%
-independent)\ and traceless operator belonging to the space of linear
operators acting on the system Hilbert space. Thus from Eq.~(\ref{H_1}) we
obtain 
\begin{equation}
H_{1}=H_{A}.
\end{equation}

Note that, from Eqs.~(\ref{H_0}) and (\ref{V_0}), 
\begin{equation}
V_{0}(\tau )=e^{-iTH_{T}(0)\int^{\tau }\Delta },  \label{V_0-form}
\end{equation}%
where $\int^{\tau }\Delta $ is 
shorthand for $\int_{0}^{\tau }\Delta (\tau
^{\prime })~\mathrm{d}\tau ^{\prime }$---we shall use the similar shorthand 
\begin{equation}
\int^{\tau }g\equiv \int_{0}^{\tau }g(\tau ^{\prime })~\mathrm{d}\tau
^{\prime }
\end{equation}%
wherever convenient. Inserting $V_{0}$ into Eq.~(\ref{V_1}) yields 
\begin{equation}
i\dot{V}_{1}=\mathcal{K}_{T}V_{1},  \label{V_1-eq}
\end{equation}%
in which the kernel $\mathcal{K}_{T}$ is defined as 
\begin{equation}
\mathcal{K}_{T}(\tau )\equiv ih_{A}(\tau )~e^{iTH_{T}(0)\int^{\tau }\Delta
}~\Xi ~e^{-iTH_{T}(0)\int^{\tau }\Delta }.  \label{kernel}
\end{equation}%
Equation (\ref{V_1-eq}), or equivalently the Volterra equation 
\begin{equation}
\ V_{1}(\tau )=\openone+\int_{0}^{\tau }\mathcal{K}_{T}(\tau ^{\prime
})V_{1}(\tau ^{\prime })~\mathrm{d}\tau ^{\prime },
\end{equation}%
can be solved iteratively, yielding the Dyson series 
\begin{eqnarray}
V_{1}(\tau ) &=&\openone+\sum_{l=1}^{\infty }\int_{0}^{\tau }\mathcal{K}%
_{T}(\tau _{1})\mathrm{d}\tau _{1}\ldots \int_{0}^{\tau _{l-1}}\mathcal{K}%
_{T}(\tau _{l})\mathrm{d}\tau _{l}.  \notag \\
&&
\end{eqnarray}


\subsubsection{Quantum search}

Now we apply the method described above to the adiabatic quantum search
problem. Recall that we are working with the reduced states and operators
(hence the hat over all reduced quantities, except the gap $\Delta $ [Eq.~(%
\ref{gap})]). Comparing Eq.~(\ref{Hhat}) with Eq.~(\ref{H_T}) implies that 
\begin{equation}
\widehat{H}(0)/J=\frac{1}{2}\sigma _{z},
\end{equation}%
after excluding the trivial term $\propto \openone$ from $\widehat{H}$.
Similarly, Eqs.~(\ref{formofA}) and (\ref{H_A}) yield 
\begin{eqnarray}
\widehat{h}_{A} &=&\dfrac{\mathrm{d}}{\mathrm{d}\tau }\left[ \dfrac{1}{2}%
\arccos [(x_{1}-(1-2r)x_{2})/\Delta ]\right]  \notag \\
&\overset{\text{(\ref{gap})}}{=}&\sqrt{r(1-r)}\frac{x_{1}\dot{x}_{2}-\dot{x}%
_{1}x_{2}}{\Delta ^{2}},  \label{h_A} \\
\widehat{\Xi } &=&\sigma _{y}.
\end{eqnarray}%
As a result, from Eqs.~(\ref{V_0-form}) and (\ref{kernel}) we obtain 
\begin{eqnarray}
\widehat{V}_{0} &=&e^{-i\widehat{\chi }}\mathrm{diag}\left( e^{-\frac{i}{2}%
JT\int^{\tau }\Delta },e^{\frac{i}{2}JT\int^{\tau }\Delta }\right) ,
\label{v0} \\
\widehat{\mathcal{K}}_{T} &=&\widehat{h}_{A}\left( 
\begin{array}{cc}
0 & e^{iJT\int^{\tau }\Delta } \\ 
-e^{-iJT\int^{\tau }\Delta } & 0%
\end{array}%
\right) ,
\end{eqnarray}%
where the phase factor 
\begin{equation}
\widehat{\chi }(\tau )=\frac{1}{2}JT\int_{0}^{\tau }[x_{1}(\tau ^{\prime
})+x_{2}(\tau ^{\prime })]~\mathrm{d}\tau ^{\prime }
\end{equation}%
compensates for the removal of the trivial term from $\widehat{H}_{T}(0)$.
Some simple algebra then yields 
\begin{equation}
\widehat{V}_{1}(\tau )=\sum_{l=0}^{\infty }(-1)^{l}\left( 
\begin{array}{cc}
\mathcal{I}_{2l}(\tau ) & \mathcal{I}_{2l+1}^{\ast }(\tau ) \\ 
-\mathcal{I}_{2l+1}(\tau ) & \mathcal{I}_{2l}^{\ast }(\tau )%
\end{array}%
\right) ,  \label{v1}
\end{equation}%
where, for $l\geq 1$, the Dyson series terms are 
\begin{equation}
\mathcal{I}_{l}(\tau )\equiv \int_{0}^{\tau }\widehat{h}_{A}(\tau ^{\prime })%
\mathcal{I}_{l-1}(\tau ^{\prime })e^{i(-1)^{l}JT\int^{\tau ^{\prime }}\Delta
}\mathrm{d}\tau ^{\prime },  \label{IntTran}
\end{equation}%
and $\mathcal{I}_{0}(\tau )\equiv 1$. This completes the derivation of 
\begin{equation}
\widehat{\Omega }_{T}=\widehat{A}^{\dag }\widehat{V}=\widehat{V}_{0}\widehat{%
V}_{1}.  \label{o-hat}
\end{equation}


\section{Adiabatic error in the search algorithm}

\label{AdError}

In the previous section, we worked out the solution to the Schr\"{o}dinger
equation in the quantum search problem. Having collected the pertinent
ingredients, we now return to calculating our main object of interest, $%
\delta _{\text{ad}}(\tau )$.

After deriving an exact formula for the error, we proceed with approximating
it in the large system-size limit, identified with $r\ll 1$. We start with
the well-known polynomial expansion of $\delta _{\text{ad}}(1)$ in terms of $%
T$, which works well for large times. In refining this result, we show that
in fact two regimes are discernible in the behavior of $\delta _{\text{ad}%
}(1)$ vs $T$: (i) the onset of \textit{exponential} decrease, followed by
(ii) a \textit{polynomial} tail. This dichotomy will appear to be crucial in
a correct characterization of the scaling of the run time of the algorithm
with system size.

The exact behavior of the error in the algorithm depends strongly on the
form of the interpolation one chooses for the Hamiltonian. Inspired by our
earlier study aiming at minimizing the adiabatic error in quantum algorithms 
\cite{AQC-intrinsic}, we shall suggest a general class of interpolations,
which includes three specific cases already studied in the literature. Next,
we investigate the specific behavior of the adiabatic error for each
interpolation, separately. Finally, we shall suggest methods for suppressing
the adiabatic error even further.


\subsection{Exact relation}

\label{exactr}

Recall that the adiabatic quantum search Hamiltonian (\ref{search-Ham}) has
a nondegenerate ground state $|\phi \rangle $ at the initial time $\tau =0$,
whereas the ground-state eigenprojection at the final time $\tau =1$ is $P_{%
\mathcal{M}}$, which is $M$-fold degenerate. Here, any full superposition of
the form $|\psi _{T}(1)\rangle =\sum_{i\in \mathcal{M}}\psi _{i}|i\rangle $
will work equally well, whereas if $|\psi _{T}(1)\rangle $ does not have
complete support over $P_{\mathcal{M}}$ then this indicates that the
algorithm has 
partially failed. Hence, following the discussion in Sec.~\ref%
{Ad-Int} [Eq.~(\ref{error-final-form})], the adiabatic error at the final
time is determined by 
\begin{eqnarray}
\delta _{\text{ad}}^{\prime }(1) &=&\sqrt{1-\langle \psi _{T}(1)|P_{\mathcal{%
M}}|\psi _{T}(1)\rangle }  \notag \\
&=&\sqrt{1-M|\psi _{\text{m}}(1)|^{2}}  \notag \\
&\overset{\text{(\ref{normalization})}}{=}&\sqrt{N}\sqrt{1-r}|\psi _{\text{u}%
}(1)|.  \label{er}
\end{eqnarray}

An equivalent formulation can be obtained for the two-dimensional reduction
we discussed in Sec.~\ref{QSHam}. In this representation $|\psi
_{T}(1)\rangle $ is replaced by $|\widehat{\psi }_{T}(1)\rangle =\widehat{V}%
_{T}(1)|\widehat{\Phi }_{-}(0)\rangle $ [Eq.~(\ref{state-red})]; similarly,
the instantaneous ground state is represented by the nondegenerate state $|%
\widehat{\Phi }_{-}(1)=\widehat{A}(1)|\widehat{\Phi }_{-}(0)\rangle $ [Eq.~(%
\ref{phi-hat})]. Thus we can employ the error formula appropriate for the
non-degenerate 
case [Eq.~(\ref{def:ad-err2})], whereby 
\begin{eqnarray}
\delta _{\text{ad}}(1) &=&\sqrt{1-|\langle \widehat{\Phi }_{-}(1)|\widehat{%
\psi }_{T}(1)\rangle |^{2}}  \notag \\
&=&\sqrt{1-|\langle z,-|\widehat{\Omega }_{T}(1)|z,-\rangle |^{2}}  \notag
\label{ad-error1} \\
&=&|\langle z,+|\widehat{\Omega }_{T}(1)|z,-\rangle |=|\langle \widehat{\Phi 
}_{+}(1)|\widehat{\psi }_{T}(1)\rangle |,  \label{ad-error}
\end{eqnarray}%
in which $\widehat{\Omega }_{T}(1)$ is given by Eq.~(\ref{o-hat}), and in
the last line we used the unitarity of $\widehat{\Omega }_{T}(1)$. The
equality of Eqs.~(\ref{er}) and (\ref{ad-error}) is immediately seen from 
\begin{eqnarray}
|\langle \widehat{\Phi }_{+}(1)|\widehat{\psi }_{T}(1)\rangle | &\overset{%
\text{(\ref{psi-hat})}}{=}&\sqrt{M}\Bigl|(%
\begin{smallmatrix}
1 & 0%
\end{smallmatrix}%
)\widehat{A}^{\dag }(1)S^{-1}\left( 
\begin{smallmatrix}
\psi _{\text{u}}(1) \\ 
\psi _{\text{m}}(1)%
\end{smallmatrix}%
\right) \Bigr|  \notag \\
&\overset{\text{(\ref{formofS}),(\ref{formofA})}}{=}&\sqrt{N}\sqrt{1-r}|\psi
_{\text{u}}(1)|.
\end{eqnarray}

Inserting $\widehat{\Omega }_{T}(1)$---noting Eqs.~(\ref{v0}) and (\ref{v1}%
)---into Eq.~(\ref{ad-error}) yields the following \textit{exact} expression
for the adiabatic error at the final time: 
\begin{equation}
\delta _{\text{ad}}(1)=\Bigl|\sum_{l=0}^{\infty }(-1)^{l}\mathcal{I}%
_{2l+1}(1)\Bigr|,  \label{error}
\end{equation}%
which is upperbounded by 
\begin{equation}
\delta _{\text{ad}}(1)\leq \sum_{l=0}^{\infty }\bigl|\mathcal{I}_{2l+1}(1)%
\bigr|.  \label{error'}
\end{equation}%
From the above equations we can in principle calculate the adiabatic error
or its bound given that we know all $\mathcal{I}_{\text{odd}}(1)$'s.

\textit{Remark.} The value of $\delta _{\text{ad}}(T=0)$ will be important
later. In this case, we have $|\widehat{\psi }_{T=0}(1)\rangle =|\Phi
_{-}(0)\rangle $, thence Eq.~(\ref{ad-error}) yields 
\begin{equation}
\delta _{\text{ad}}(0)=\sqrt{1-r},
\label{T=0-eq}
\end{equation}%
in which we used the boundary conditions (\ref{BC-2}) in $\widehat{A}(1)$.
This relation is valid for any interpolation that satisfies the boundary
conditions.


\subsection{Approximation of the adiabatic error}

\label{Approx}

An exact calculation of the adiabatic error from Eq.~(\ref{error}) can be
challenging because of the infinite number of terms in the summand and the
fact that each term contains a multiple integral. To alleviate this
difficulty, in this subsection we approximate the upper bound on $\delta _{%
\text{ad}}(1)$ from Eq.~(\ref{error'}) from the first few $\mathcal{I}_{l}$%
's, and argue that this suffices for most algorithmic purposes. We start
from an expansion in powers of $1/T$, based on integration by parts, and
explain its limitations. We then provide more careful analyses, based on the
residue theorem and on the stationary phase method, both of which 
lead to an exponential error estimate.


\subsubsection{Polynomial expansion}

The most common rigorous adiabatic approximation employs an expansion in
powers of $1/T$, presuming that $T$ is \textquotedblleft
large\textquotedblright\ \cite{Avron:871,JRS,LRH}. Let us now show how one
can systematically expand the adiabatic error as a polynomial in$1/T$ by
extracting powers of $1/T$ from each term $\mathcal{I}_{l}(\tau )$ through
integration by parts.

From the identity $e^{-iTY(\tau )}=
i/\bigl(T\dot{Y}(\tau )\bigr)\frac{\mathrm{d}%
}{\mathrm{d}\tau }e^{-iTY(\tau )}$ [valid for any differentiable function $%
Y(\tau )$], we obtain the following relation by integration by parts: 
\begin{eqnarray}
&&\int_{0}^{\tau }G(\tau ^{\prime })e^{-iTY(\tau ^{\prime })}\mathrm{d}\tau
^{\prime }=  \label{ibp-identity} \\
&&\frac{i}{T}\Bigl[e^{-iTY(\tau ^{\prime })}\frac{G(\tau ^{\prime })}{\dot{Y}%
(\tau ^{\prime })}\Big|_{0}^{\tau }-\int_{0}^{\tau }e^{-iTy(\tau ^{\prime })}%
\frac{\mathrm{d}}{\mathrm{d}\tau ^{\prime }}\Bigl(\frac{G(\tau ^{\prime })}{%
\dot{Y}(\tau ^{\prime })}\Bigr)\mathrm{d}\tau ^{\prime }\Bigr].  \notag
\end{eqnarray}%
Notice how this extracted a $1/T$ in front of the first term. In the second
integral on the right hand side we can iterate the same trick of replacing
the exponential with its derivative; which generates $1/T^{2}$ and higher
order terms. This provides a systematic way for generating $\mathrm{poly}%
(1/T)$ expansions of exponential integrals, as we shall see more
specifically below for the $\mathcal{I}_{l}(\tau )$'s.

Using Eq.~(\ref{ibp-identity}), we obtain 
\begin{eqnarray}
\mathcal{I}_{1}(\tau ) &=&\frac{i}{JT}\Bigl[e^{-iJT\int^{\tau ^{\prime
}}\Delta }\frac{\widehat{h}_{A}(\tau ^{\prime })}{\Delta (\tau ^{\prime })}%
\Bigl|_{0}^{\tau }-e^{-iJT\int^{\tau }\Delta }  \notag \\
&&\times \int_{0}^{\tau }\frac{\mathrm{d}}{\mathrm{d}\tau ^{\prime }}\Bigl(%
\frac{\widehat{h}_{A}(\tau ^{\prime })}{\Delta (\tau ^{\prime })}\Bigr)~%
\mathrm{d}\tau ^{\prime }\Bigr].  \label{i1}
\end{eqnarray}%
Applying once more the exponential identity (\ref{ibp-identity}) for the
second integral above gives rise to $O(1/T^{2})$ terms; whence, 
\begin{eqnarray}
\mathcal{I}_{1}(\tau ) &=&\frac{i}{JT}\Bigl[\frac{\widehat{h}_{A}(\tau )}{%
\Delta (\tau )}e^{-iJT\int^{\tau }\Delta }-\widehat{h}_{A}(0)\Bigr]+O\Bigl(%
\frac{1}{T^{2}}\Bigr),  \notag \\
&&
\end{eqnarray}%
i.e., $|\mathcal{I}_{1}(\tau )|=O(1/T)$. In analogous fashion, for $\mathcal{%
I}_{2}(\tau )$ we obtain 
\begin{eqnarray}
\mathcal{I}_{2}(\tau ) &=&\frac{i}{JT}\Bigl[\int_{0}^{\tau }\frac{\widehat{h}%
_{A}^{2}(\tau ^{\prime })}{\Delta (\tau ^{\prime })}\mathrm{d}\tau ^{\prime
}-\widehat{h}_{A}(0)\mathcal{I}_{1}^{\ast }(\tau )  \notag \\
&&-\int_{0}^{\tau }\mathrm{d}\tau ^{\prime }\widehat{h}_{A}(\tau ^{\prime
})e^{iJT\int^{\tau ^{\prime }}\Delta }  \notag \\
&&\times \int_{0}^{\tau ^{\prime }}\mathrm{d}\tau ^{\prime \prime }\frac{%
\mathrm{d}}{\mathrm{d}\tau ^{\prime \prime }}\Bigl(\frac{\widehat{h}%
_{A}(\tau ^{\prime \prime })}{\Delta (\tau ^{\prime \prime })}\Bigr)%
e^{-iJT\int^{\tau ^{\prime \prime }}\Delta }\Bigr],  \notag \\
&&
\end{eqnarray}%
from which 
\begin{eqnarray}
\mathcal{I}_{2}(\tau ) &=&\frac{i}{JT}\Bigl[\int_{0}^{\tau }\frac{\widehat{h}%
_{A}^{2}(\tau ^{\prime })}{\Delta (\tau ^{\prime })}\mathrm{d}\tau ^{\prime }%
\Bigr]-\frac{\widehat{h}_{A}(0)}{(JT)^{2}}\Bigl[\frac{\widehat{h}_{A}(\tau
)e^{-iJT\int^{\tau }\Delta }}{\Delta (\tau )}  \notag \\
&&-\widehat{h}_{A}(0)\Bigr]+O\Bigl(\frac{1}{T^{3}}\Bigr),  \label{i2}
\end{eqnarray}%
and $|\mathcal{I}_{2}(\tau )|=O(1/T)$. By induction, one can conclude from
Eq.~(\ref{IntTran}) that \cite{Avron:871,Avron:872} 
\begin{equation}
|\mathcal{I}_{2l-1}(\tau )|=|\mathcal{I}_{2l}(\tau )|=O\Bigl(\frac{1}{T^{l}}%
\Bigr),  \label{OI}
\end{equation}%
for $l\in \mathbb{N}$. Thus, from Eq.~(\ref{error'}) the adiabatic error
bound becomes 
\begin{eqnarray}
\delta _{\text{ad}}(1) &\leq &|\mathcal{I}_{1}(1)|+O\Bigl(\frac{1}{T^{2}}%
\Bigr)  \notag \\
&=&\frac{1}{JT}\Bigl[|\widehat{h}_{A}(0)|+|\widehat{h}_{A}(1)|\Bigr]+O\Bigl(%
\frac{1}{T^{2}}\Bigr).  \label{addelta}
\end{eqnarray}%
This relation can be simplified further. From the boundary conditions (\ref%
{BC-1}) and (\ref{BC-2}), we obtain $\Delta (0)=\Delta (1)=1$ [Eq.~(\ref{gap}%
)], $\widehat{h}_{A}(0)=\sqrt{r(1-r)}\dot{x}_{2}(0)$ and $\widehat{h}%
_{A}(1)=-\sqrt{r(1-r)}\dot{x}_{1}(1)$ [Eq.~(\ref{h_A})]. Thus Eq.~(\ref%
{addelta}) reduces to 
\begin{equation}
\delta _{\text{ad}}(1)\leq \frac{\sqrt{r(1-r)}}{JT}\Bigl[|\dot{x}_{2}(0)|+|%
\dot{x}_{1}(1)|\Bigr]+O\Bigl(\frac{1}{T^{2}}\Bigr).  \label{er-simp}
\end{equation}%
Proceeding in a similar manner, one can in principle obtain the exact form
of the coefficient of each $1/T^{l}$ term, for arbitrary $l\in \mathbb{N}$.

\textit{Remarks.} Let us make some remarks regarding the polynomial
expansion, and in particular Eq.~(\ref{er-simp}).

(i) Notice that often the $\mathrm{poly}(1/T)$ series is \emph{truncated}
after the first or at most the second order term, on the basis of the
assumption that for sufficiently large $T$ the first couple of terms should
give a reliable and accurate upper bound. However, without correctly
defining what \textquotedblleft large\textquotedblright\ $T$ means, a
truncation after the first few terms might be unjustifiable. In fact, in
addition to $T$, the system size $\log N$ (introduced here through $r$) and
the gap $\Delta (\tau )$ are also key players in the estimation of $\delta _{%
\text{ad}}(\tau )$. The minimum (system-size dependent) gap $\Delta _{\min
}\equiv \min_{\tau }\Delta (\tau )$ works in general as a bottleneck for the
performance of quantum algorithms (e.g., Refs.~\cite%
{Farhi1,Farhi2,Latorre,QAB,Schaller}). Specifically, where the gap closes or
becomes small, the adiabatic approximation may not hold, indicative of a
\textquotedblleft quantum phase transition\textquotedblright\ (in the
thermodynamic limit) \cite%
{Sachdev:book,Latorre,Schutzhold:06,Schaller2,AminChoi}. This implies that
the coefficients of some high order $1/T^{l}$ terms might have a stronger
gap dependence than those of lower order terms. With this caveat, neglecting
those higher order terms is not always possible. In fact, it is not
difficult to see that the coefficient of the $1/T^{2}$ has a $\Delta ^{-6}$
dependence (see also Ref.~\cite{JRS}), stronger than the $\Delta ^{-2}$
dependence of the coefficient of the $1/T$ term in $\delta _{\text{ad}}(\tau
)$ [Eq.~(\ref{h_A})].

(ii) For similar reasons, an estimate of $T$ arising from $\delta _{\text{ad}%
}(1)\leq \varepsilon $ (for a given $\varepsilon $) along with a truncated $%
\mathrm{poly}(1/T)$ expansion, is not always reliable. We shall see this
explicitly later in this section.

(iii) One might argue that $\delta _{\text{ad}}(1)=O(1/T)$
results from an energy-time uncertainty relation such as $\delta _{\text{ad}%
}(1)\times T\approx 1$ [presuming $\delta _{\text{ad}}(1)$ is directly
related to the uncertainty in measurement of energy]. However, this argument
is not rigorous and should not be considered as a replacement for the
analysis leading to the $\mathrm{poly}(1/T)$ expansion
(unless justified carefully).  
A rigorous energy-time uncertainty relation is given, for example, by the Mandelstam-Tamm inequality
\begin{equation}
\Delta _{\psi }[H]\times T_{\psi }[K]\geq 1/2,
\end{equation}%
in which $T_{\psi }[K]\equiv \Delta _{\psi }[K]/|\mathrm{d}\langle \psi
|K|\psi \rangle /\mathrm{d}t|$, $K$ is any observable, and $\Delta _{\psi
}[X]\equiv \sqrt{\langle \psi |X^{2}|\psi \rangle -\langle \psi |X|\psi
\rangle ^{2}}$ \cite{Galindo1:book}. Hence, as is well known the naive
relation $\Delta _{\psi }[H]\times T\approx 1$ should be treated with care.
Strictly, a relation between $\Delta _{\widehat{\psi }_{T}(1)}[\widehat{H}%
(1)]$ and $\delta _{\text{ad}}(1)$ can be constructed as the following. Note
that we can write 
\begin{equation}
|\widehat{\psi }_{T}(\tau )\rangle =\sqrt{1-\delta _{\text{ad}}(\tau )}|%
\widehat{\Phi }_{-}(\tau )\rangle +\delta _{\text{ad}}(\tau )|\widehat{\Phi }%
_{-}^{\perp }(\tau )\rangle ,
\end{equation}%
where $|\widehat{\Phi }_{-}^{\perp }(\tau )\rangle $ is orthogonal to $|%
\widehat{\Phi }_{-}(\tau )\rangle $ [Eq.~\ref{ad-error}]. Hence, after some
algebra we obtain 
\begin{eqnarray}
\hskip-3mm &&\Delta _{\widehat{\psi }_{T}(1)}\widehat{H}(1)\approx \delta _{%
\text{ad}}(1)  \notag \\
\hskip-3mm &&~\times \sqrt{2\widehat{E}_{-}\bigl[\widehat{E}_{-}\langle 
\widehat{\Phi }_{-}^{\perp }|H^{2}|\widehat{\Phi }_{-}^{\perp }\rangle
-\langle \widehat{\Phi }_{-}^{\perp }|H|\widehat{\Phi }_{-}^{\perp }\rangle %
\bigr]\bigl|_{\tau =1}}+O(\delta _{\text{ad}}^{2}).  \notag \\
\hskip-3mm &&
\end{eqnarray}%
Despite this relation, connecting $T$ and $T_{\widehat{\psi }_{T}(1)}[K]$ 
is not straightforward. Although using different versions of the energy-time uncertainty relation \cite{ETU-1,ETU-2,ETU-3} 
may provide additional insights, we shall not further pursue this here.


\subsubsection{Exponential estimate}

\label{expl}

\paragraph{Residue theorem analysis}

In the previous subsection we used integration by parts to arrive at a
polynomial expansion. Let us now show that an alternative, more careful
analysis based on the residue theorem of complex analysis, reveals that the
adiabatic error decays \textit{exponentially} for sufficiently short times.
In some sense, this exponential behavior is reminiscent of the well-known
Landau-Zener formula for two-state quantum systems \cite{LZ-Avron},
which---in its simplest form---states that the \textit{tunneling probability}
$p_{T}(1)\equiv |\langle \psi (0)|V_{T}^{\dag }(1)|\Phi _{1}(1)\rangle |^{2}$
from $|\psi (0)\rangle =|\Phi _{0}(0)\rangle $ to $|\Phi _{1}(1)\rangle $,
for the Hamiltonian $H(\tau )/J=(\tau \sigma _{z}+D\sigma _{x})/2$, is 
\begin{equation}
p_{T}(1)=e^{-\pi JTD^{2}/2}.
\end{equation}%
We notice that $p_{T}(1)$ is in fact intimately related to $\delta _{\text{ad%
}}(1)$ in this simple two-state case, as 
\begin{equation}
p_{T}(1)=1-\delta _{\text{ad}}^{2}(1).
\end{equation}%
The exponentiality of the adiabatic error \textit{vs} time and (some power
of) the gap has been previously established in generality in rigorous
treatments of the adiabatic theorem \cite%
{Nenciu:93,Martinez,HagedornJoye:02,LRH} (see also Ref. \cite{Schaller} in a
more restricted setting).

In what follows we focus on $\mathcal{I}_{1}(\tau )$; higher order terms can
be treated similarly. We start from Eq.~(\ref{error'}), whence 
\begin{equation}
\delta _{\text{ad}}(1)\leq \left\vert \int_{0}^{1}\widehat{h}_{A}(\tau
)e^{iJT\int^{\tau }\Delta }\mathrm{d}\tau \right\vert +\ldots  \label{h1-int}
\end{equation}%
Here \textquotedblleft \ldots \textquotedblright\ denotes the higher order
terms $|\mathcal{I}_{2l+1}(1)|$ ($l\geq 1$), whose neglect we justify below
in the specific examples we discuss. An exponential error term can be
obtained, for example, by extending the integral to the complex time plane
and using an appropriate closed contour for the application of the residue
theorem \cite{Arfken:book}. A precursor to this treatment of the adiabatic
error can be found, e.g., in Ref.~\cite{Schaller}.

If $\widehat{h}_{A}(z)$ ($z\in \mathbb{C}$) is not constant it has poles at
points $z_{0}$ where the gap vanishes: $\Delta (z_{0})=0$ [Eq.~(\ref{h_A})].
From Eq.~(\ref{gap}), we obtain $x_{1}(z_{0})=x_{2}(z_{0})=0$, or (assuming $%
x_{1}(z_{0})\neq 0$) 
\begin{equation}
x_{21}({z_{0}}_{\pm })=1-2r\pm 2i\sqrt{r(1-r)},  \label{poles}
\end{equation}%
where $x_{21}\equiv x_{2}/x_{1}$. The poles $z_{0}$ can in principle be
obtained by inverting this relation for a given interpolation $x_{21}(\tau )$%
. Note, however, that there might exist other singularities (typically at
infinity) arising from the exponential $e^{-iJT\int^{z}\Delta }$ in Eq. (\ref%
{h1-int}). Therefore, estimating the integral (\ref{h1-int}) requires
finding all contributing singularities in a suitably chosen contour $%
\mathcal{C}$ in the complex $\tau $-plane.

The value of the integral $|\int_{0}^{1}\widehat{h}_{A}(\tau
)e^{iJT\int^{\tau }\Delta }\mathrm{d}\tau |$ can now be obtained by
calculating the residues of the integrand at the poles enclosed inside the
contour, 
\begin{eqnarray}
|\mathcal{I}_{1}(1)| &=&\Bigl|2\pi i\sum_{z_{0}\in \text{inside}(\mathcal{C}%
)}\mathrm{Res}\Bigl[\widehat{h}_{A}(z)e^{iJT\int^{z}\Delta },z_{0}\Bigr] 
\notag \\
&&~-\int_{\mathcal{C}-[0,1]}\widehat{h}_{A}(z)e^{iJT\int^{z}\Delta }\mathrm{d%
}z\Bigr|.
\label{tmp-eq-1}
\end{eqnarray}%
For example, if the poles $z_{0}$ are \textit{simple} and arise from $\Delta
(z_{0})=0$, calculating the integral becomes straightforward. Recall that $%
\widehat{h}_{A}\propto 1/\Delta ^{2}$ [Eq.~(\ref{h_A})] and the residue of a
rational function $P(z)/Q(z)$ at a simple pole $z^{\star }$ is given by $%
P(z^{\star })/\partial_z Q(z^{\star })$
\cite{Arfken:book}. Thus, in this case we obtain 
\begin{eqnarray}
\hskip-5mm &&\mathrm{Res}\Bigl[\widehat{h}_{A}(z)e^{iJT\int^{z}\Delta },z_{0}%
\Bigr]=  \label{residue} \\
\hskip-5mm &&\frac{\sqrt{r(1-r)}\dot{x}_{21}(z_{0})e^{iJT\int^{z_{0}}\Delta }%
}{\frac{\mathrm{d}}{\mathrm{d}z}\bigl[\bigl(1-x_{21}(z)\bigr)^{2}+4rx_{21}(z)%
\bigr]\bigl|_{z_{0}}}\approx
\frac{1}{4i}e^{iJT\int^{z_{0}}\Delta } \notag,
\end{eqnarray}%
where in the last line we assumed $\dot{x}_{21}(z_{0})\neq 0$. 
Note that this computation of the residues does not necessarily hold
when $\widehat{h}_{A}=\mathrm{const.}$, or when $z_{0}$ is at infinity.

This yields that an exponentially decreasing contribution emerges from $%
\mathrm{Im}\bigl[\int^{z_{0}}\Delta \bigr]$ of the exponential within the
integrand---perhaps in addition to a generically
non-exponentially-decreasing term. Thus, from Eq.~(\ref{residue}) we find 
\begin{equation}
\hskip-1mm\delta _{\text{ad}}(1)\lessapprox \sum_{z_{0}\in \text{inside}(%
\mathcal{C})}R_{T}(z_{0})~e^{-JT~\mathrm{Im}\left[ \int^{z_{0}}\Delta \right]
}+R_{T}^{\prime }(1),  \label{xx}
\end{equation}%
where $R_{T}(z_{0})\in \mathbb{R}^{+}$ results from the non-exponential
contribution of the residue at $z_{0}$ and $R_{T}^{\prime }(1)\in
\mathbb{R}^{+} $ encapsulates the collective nonvanishing contribution
of other segments of the contour as well as other non-exponentially-decreasing
contributions emerging from the \textquotedblleft \ldots \textquotedblright\
terms in Eq.~(\ref{h1-int}).\footnote{In the case of
Eq.~(\ref{residue}), and from Eq.~(\ref{tmp-eq-1}), we obtain:
$R_T=\pi/2$. This is in good agreement with Eq.~(\ref{T=0-eq}) when
$T=0$ and $r\ll1$.}

\paragraph{Stationary phase analysis}

An alternative and complementary approach for obtaining the exponential
contribution to the adiabatic error is to use the stationary phase method.
This method is appropriate for obtaining asymptotic value of complex
integrals of the form $\int_{\gamma }F(z)e^{iTG(z)}\mathrm{d}z$, where $T>0$
is a large number and $\gamma $ is a path for the integration. Assuming $F(z)
$ is a slowly-varying function over $\gamma $ and $G(z)$ is an analytic
function, one can argue that the major contribution to the integral comes
from the point(s) $z_{0}$ at which $G(z)$ has a minimum, whence \cite%
{Hayek:book} 
\begin{equation}
\int_{\gamma }F(z)e^{iTG(z)}\mathrm{d}z\approx \sqrt{\frac{2\pi }{T\ddot{G}%
(z_{0})}}F(z_{0})e^{i\pi /4}e^{iTG(z_{0})}.  \label{spm}
\end{equation}

To apply this result to Eq.~(\ref{h1-int}), we replace $G(z)$ and $F(z)$
with $J\int^{z}\Delta $ and $\widehat{h}_{A}(z)$, respectively. Accordingly, 
$z_{0}$ is where $\dot{G}(z_{0})=\Delta (z_{0})=0$. The minimum point $z_{0}$
is often a complex number. In fact, in quantum many-body situations, the gap 
$\Delta $ is a nonnegative function, often with a nonvanishing minimum,
which becomes zero in the thermodynamic limit \cite{Sachdev:book}. In such
cases, $G(z)$ does indeed have a minimum. It is also required that $%
\widehat{h}_{A}(z_{0})$ be finite [this of course is not satisfied when $%
\widehat{h}_{A}$ has a pole or singularity at $z_{0}$]. If all these
conditions are satisfied, Eqs.~(\ref{h1-int}) and (\ref{spm}) yield 
\begin{equation}
\delta _{\text{ad}}(1)\lessapprox \left\vert \widehat{h}_{A}(z_{0})\sqrt{%
\frac{2\pi }{JT\dot{\Delta}(z_{0})}}\right\vert e^{-JT\mathrm{Im}%
[\int^{z_{0}}\Delta ]},  \label{st-ph}
\end{equation}%
as we wished. This relation complements Eq.~(\ref{residue}) in that it may
be applicable when Eq.~(\ref{residue}) is not.

\paragraph{Discussion}

The emergence of the exponential dependence of $\delta _{\text{ad}}$ on $T$
[Eqs.~(\ref{xx}) and (\ref{st-ph})] in contrast to the polynomial dependence
[Eq.~(\ref{er-simp})] is remarkable, as it indicates a much faster decay of
the adiabatic error than what is suggested by the standard \thinspace $1/T$
expansion. In the exponential regime it suffices that $T$ be large compared
to $1/\mathrm{Im}[\int^{z_{0}}\Delta ]$, or roughly \cite{Schaller}:%
\begin{equation}
T\gg \frac{1}{J\Delta }.
\end{equation}%
This is a less stringent condition than the standard condition (\ref{cond})
or its more rigorous counterparts \cite%
{Avron:871,Avron:872,Nenciu:93,Martinez,HagedornJoye:02,JRS,o'hara:042319,Boixo,LRH}%
, involving higher powers of the gap.\ The crossover point $T^{\star }$
between the exponential and the polynomial regimes can be estimated by
solving 
\begin{equation}
R_{T^{\star }}(r)~e^{-JT^{\star }\mathrm{Im}[\int^{z_{0}}\Delta ]}\approx 
\frac{\sqrt{r}}{JT^{\star }}\bigl(|\dot{x}_{1}(1)|+|\dot{x}_{2}(0)|\bigr),
\label{crossover}
\end{equation}%
in which $R_{T}(r)$ is a non-exponential prefactor given by Eq.~(\ref{xx})
or (\ref{st-ph}). If $T\lesssim T^{\star }$ ($T\gtrsim T^{\star }$) the
exponential (polynomial) behavior prevails.

Later in this section, we show explicitly that the expected run time for the
quantum search algorithm is often given by the exponential contribution; the
polynomial regime may \emph{overestimate} the minimum run time required for
reasonable accuracy.

Since the scaling of the run time depends on which interpolation we choose
for the Hamiltonian, in the following we shall obtain specific
interpolations by employing some recently developed results for (partial)
minimization of the adiabatic error \cite{QAB,AQC-intrinsic}.


\subsection{Hamiltonian interpolation}

\label{HAM}

The set of available control knobs ($\mathbf{x}$), as well as the way one
varies them, determine the specific Hamiltonian interpolation implemented in
a laboratory setting. Theoretically, though, there are various ways by which
one can obtain families of Hamiltonians for an adiabatic quantum algorithm.
One natural choice is interpolations which minimize \textquotedblleft
physical\textquotedblright\ cost. For example, in the setting of Refs.~\cite%
{RolandCerf,QAB}, time functionals were constructed from a \emph{local}
version of the adiabatic condition (\ref{cond}), whose minimization resulted
in a set of Euler-Lagrange equations for the underlying interpolations. A
different method was suggested in Ref.~\cite{AQC-intrinsic}, where it was
shown that in the standard $\mathrm{poly}(1/T)$ expansion of the adiabatic
error $\delta _{\text{ad}}(\tau )$ [general counterpart of Eq.~(\ref{addelta}%
) or (\ref{er-simp})], the coefficient of the $1/T$ term has a \emph{%
geometric} part, in a differential geometric sense. Specifically, this
geometric coefficient, in terms of the ground-state eigenprojection, $%
P(\tau )$ is%
\begin{equation}
\int_{0}^{\tau }\bigl\Vert\lbrack \dot{P}\bigl(\mathbf{x}(\tau ^{\prime })%
\bigr),P\bigl(\mathbf{x}(\tau ^{\prime })\bigr)]\bigr\Vert~\mathrm{d}\tau
^{\prime }.  \label{geometric}
\end{equation}%
Minimizing this coefficient yields adiabatic \textquotedblleft
geodesic\textquotedblright\ interpolations, which partially decrease $\delta
_{\text{ad}}(\tau )$ for a given $T$. It follows from standard variational
calculus \cite{Arfken:book,Nakahara:book} that the geodesic interpolations
satisfy the following equation 
\begin{equation}
\bigl\Vert\lbrack \dot{P}\bigl(\mathbf{x}(\tau )\bigr),P\bigl(\mathbf{x}%
(\tau )\bigr)]\bigr\Vert\Bigl|_{\mathbf{x}_{\text{geo}}(\tau )}=\text{const.}%
,
\label{geo-eq}
\end{equation}%
where the constant is chosen to satisfy boundary conditions. We adopt this
method in the following and derive geodesic interpolations for the adiabatic
quantum search.


\subsubsection{General case}

\label{path-g}

In the reduced two-dimensional representation, $P$ is replaced with $%
\widehat{P}_{-}=|\widehat{\Phi }_{-}\rangle \langle \widehat{\Phi }_{-}|$.
Thus we have 
\begin{eqnarray}
\lbrack \dot{\widehat{P}}_{-},\widehat{P}_{-}\rbrack &=&|\widehat{\Phi }%
_{-}\rangle \langle \dot{\widehat{\Phi }}_{-}|-|\dot{\widehat{\Phi }}%
_{-}\rangle \langle \widehat{\Phi }_{-}|+|\widehat{\Phi }_{-}\rangle \langle 
\widehat{\Phi }_{-}|  \notag \\
&&\times \bigl(\langle \dot{\widehat{\Phi }}_{-}|\widehat{\Phi }_{-}\rangle
-\langle \widehat{\Phi }_{-}|\dot{\widehat{\Phi }}_{-}\rangle \bigr).
\end{eqnarray}%
Noting with the help of Eqs. (\ref{formofA}), (\ref{phi-hat}), and (\ref{h_A}%
) that 
\begin{equation}
\langle \widehat{\Phi }_{-}|\dot{\widehat{\Phi }}_{-}\rangle =i\widehat{h}%
_{A}\langle z,-|\widehat{A}^{\dag }\sigma _{y}\widehat{A}|z,-\rangle =0,
\end{equation}%
we find 
\begin{eqnarray}
\bigl\Vert\lbrack \dot{\widehat{P}}_{-},\widehat{P}_{-}]\bigr\Vert &=&\sqrt{%
2\langle \dot{\widehat{\Phi }}_{-}|\dot{\widehat{\Phi }}_{-}\rangle }  \notag
\\
&\overset{\text{(\ref{phi-hat})}}{=}&\sqrt{2}|\widehat{h}_{A}|.
\end{eqnarray}%
Hence, according to Eq.~(\ref{geo-eq}), a class of adiabatic geodesics can
be obtained from 
\begin{equation}
\widehat{h}_{A}\bigl(\mathbf{x}(\tau )\bigr)=\mathrm{const.}\equiv \varphi ,
\end{equation}%
which implies that the adiabatic Hamiltonian $\widehat{H}_{A}(\tau )$ is in
fact constant. In other words, from $\widehat{A}(\tau )=e^{-i\int^{\tau }%
\widehat{H}_{A}}$ [Eq.~(\ref{formofA})] we can see that 
\begin{equation}
\widehat{A}(\tau )=e^{-i\varphi \tau \sigma _{y}}.
\end{equation}%
This equation suggests that a more general choice for the intertwiner $%
\widehat{A}$ can be obtained by $\varphi \tau \rightarrow \varphi \theta
(\tau )$, for some arbitrary differentiable $\theta $; i.e., $\widehat{A}%
(\tau )=e^{-i\varphi \theta (\tau )\sigma _{y}}$. In terms of $\widehat{h}%
_{A}$, this translates into choosing $\widehat{h}_{A}\bigl(\mathbf{x}(\tau )%
\bigr)=\varphi \dot{\theta}(\tau )$. In terms of the geometric factor (\ref%
{geometric}), this simply means that on the geodesic, the time is determined
by $\theta (\tau )$ rather than $\tau $. To see this, note that 
\begin{eqnarray}
\int_{0}^{\tau }\bigl\Vert\lbrack \dot{P}(\tau ^{\prime }),P(\tau ^{\prime
})]\bigr\Vert~\mathrm{d}\tau ^{\prime } &=&\int_{0}^{\theta (\tau )}%
\bigl\Vert\lbrack \partial _{\theta }P(\theta ),P(\theta )]\bigr\Vert~%
\mathrm{d}\theta ,  \notag \\
&&
\end{eqnarray}%
which in turn, from variational calculus, gives the following equation for
the geodesics: 
\begin{equation}
\bigl\Vert\lbrack \partial _{\theta }P\bigl(\mathbf{x}(\theta (\tau ))\bigr)%
,P\bigl(\mathbf{x}(\theta (\tau ))\bigr)]\bigr\Vert\Bigl|_{\mathbf{x}_{\text{%
geo}}(\theta (\tau ))}=\mathrm{const.}  \label{geo-eq2}
\end{equation}%
Hence, 
\begin{eqnarray}
\bigl\Vert\lbrack \partial _{\theta }\widehat{P}_{-},\widehat{P}_{-}]%
\bigr\Vert &=&\frac{1}{\dot{\theta}}\bigl\Vert\lbrack \dot{\widehat{P}}_{-},%
\widehat{P}_{-}]\bigr\Vert  \notag \\
&=&\sqrt{2}\frac{|\widehat{h}_{A}|}{\dot{\theta}}  \notag \\
&=&\mathrm{const.},
\end{eqnarray}%
where we assumed $\dot{\theta}(\tau )>0$, $\theta (0)=0$, and $\theta (1)=1$%
. As a result, we choose the adiabatic geodesic satisfying 
\begin{equation}
\widehat{h}_{A}\Bigl(\mathbf{x}\bigl(\theta (\tau )\bigr)\Bigr)=\varphi \dot{%
\theta}(\tau ).  \label{g-path}
\end{equation}

We remark that introducing an arbitrary nondecreasing function $\theta
(\tau)$ will serve as more than a generalization. In fact, we show
below that $\theta $ allows us to substantially enhance the suppression of the adiabatic
error. For example, choosing a $\theta $ such that it is a
differentiable function of $\tau $ (up to some controllable order,
say, $k$), with vanishing derivatives (up to the same order $k$) at
the initial and final times, can suppress the adiabatic error [up to $\mathrm{poly}%
(1/T^{k+1})$]. This property---which comes at the relatively small price of
sufficient control at the beginning and end of the dynamics---may have
immediate applications in experimental realizations of quantum
annealing and AQC.

The above geodesic equation can be solved analytically. Recall that $%
x_{21}=x_{2}/x_{1}$, with the boundary values $x_{21}(0)=0$ and $%
x_{21}(1)=\infty $ [Eqs.~(\ref{BC-1}) and (\ref{BC-2})]. Then from Eq.~(\ref%
{h_A}) we find 
\begin{equation}
\frac{\sqrt{r(1-r)}\dot{x}_{21}(\tau )}{[1-x_{21}(\tau )]^{2}+4rx_{21}(\tau )%
}=\varphi \dot{\theta}(\tau ),  \label{weak}
\end{equation}%
The solution to this equation can be written as follows: 
\begin{equation}
x_{21}\bigl(\theta (\tau )\bigr)=\frac{\sin [2\theta (\tau )\varphi ]}{\sin
[2(1-\theta (\tau ))\varphi ]},  \label{relat}
\end{equation}%
where we must choose 
\begin{equation}
\varphi =\arctan \bigl[\sqrt{(1-r)/r}\bigr].
\end{equation}%
Observe that $x_{21}\bigl(\theta (\tau )\bigr)$ has the following symmetry: 
\begin{equation}
x_{21}\bigl(1-\theta (\tau )\bigr)=1/x_{21}\bigl(\theta (\tau )\bigr),
\end{equation}%
which is satisfied, for example, by requiring 
\begin{equation}
x_{2}\bigl(\theta (\tau )\bigr)=x_{1}\bigl(1-\theta (\tau )\bigr).
\label{symmetry}
\end{equation}%
Equations~(\ref{relat}) and (\ref{symmetry}) identify a two-dimensional
interpolation for the quantum search Hamiltonian.

Notice that, given Eq.~(\ref{relat}), we can also add another relation
(satisfying the boundary conditions) so as to find other classes of
interpolation. For example, we can choose 
\begin{equation}
x_{1}\bigl(\theta (\tau )\bigr)+x_{2}\bigl(\theta (\tau )\bigr)=a(\tau ),
\label{BC-3}
\end{equation}%
in which $a(\tau )$ can be a smooth function with the boundary values $%
a(0)=a(1)=1$; e.g., $a(\tau )=1+\tau (1-\tau )$ or $a(\tau )=1+\sin (\pi
\tau )$. Choosing a form for $a(\tau )$ corresponds to assuming a given $%
\Vert H(\tau )\Vert $ [Eq.~(\ref{Hnorm})], which implies a given control
over the maximum amount of the available energy in the system. Thus, Eqs.~(%
\ref{relat}) and (\ref{BC-3}) yield 
\begin{eqnarray}
\hskip-6mm x_{1}\bigl(\theta (\tau )\bigr) &=&\frac{a(\tau )\sin [2(1-\theta (\tau
))\varphi ]}{2\sin (\varphi )\cos [(1-2\theta (\tau ))\varphi ]}  \notag \\
&=&\frac{a(\tau )}{2}\Bigl[1+\cot (\varphi )\tan [(1-2\theta (\tau ))\varphi
]\Bigr],  \label{X1} \\
\hskip-6mm x_{2}\bigl(\theta (\tau )\bigr) &=&\frac{a(\tau )\sin [2\theta (\tau
)\varphi ]}{2\sin (\varphi )\cos [(1-2\theta (\tau ))\varphi ]}  \notag \\
&=&\frac{a(\tau )}{2}\Bigl[1-\cot (\varphi )\tan [(1-2\theta (\tau ))\varphi
]\Bigr].  \label{X2}
\end{eqnarray}%
Note that this interpolation also satisfies the symmetry (\ref{symmetry}).

It is clear that one can consider other auxiliary or control conditions over
the Hamiltonian different from Eq.~(\ref{BC-3}). In the following, we
address three special cases: (i) The Hamiltonian interpolation is linear in
time, namely, $x_{1}(\tau )=1-x_{2}(\tau )=1-\tau $; (ii) Hamiltonians with
constant norm; specifically, $\Vert H(\tau )\Vert /J=1$; and, (iii)
Hamiltonians with constant gap; specifically, $\Delta (\tau )=1$.


\subsubsection{Linear interpolation}

\label{lin-sec}

If we choose 
\begin{equation}
\theta (\tau )=\frac{1}{2}-\frac{1}{2\varphi }\arctan [(1-2\tau )\tan
\varphi ],
\end{equation}%
and assume $a(\tau )=1$, from Eqs.~(\ref{X1}) and (\ref{X2}) we obtain the
simple linear interpolation 
\begin{eqnarray}
&&x_{1}(\tau )=1-\tau , \\
&&x_{2}(\tau )=\tau .
\end{eqnarray}


\subsubsection{Constant-norm interpolation}

\label{c-norm}

Let us assume $0\leq x_1,x_2\leq1$. The constraint $\Vert H(\tau)\Vert/J=1$
implies that $x_1+x_2=1$ [Eq.~(\ref{Hnorm})], or equivalently $a(\tau)=1$
[Eq.~(\ref{BC-3})]. Hence, in this case from Eqs.~(\ref{X1}) and (\ref{X2})
we obtain 
\begin{eqnarray}
x_1\bigl(\theta(\tau) \bigr) =\frac{1}{2}+\frac{\sqrt{r}}{2\sqrt{1-r}}\tan%
\bigl[\bigl(1-2\theta(\tau)\bigr)\varphi\bigr],  \label{RCpath1} \\
x_2\bigl(\theta(\tau) \bigr) =\frac{1}{2}-\frac{\sqrt{r}}{2\sqrt{1-r}}\tan%
\bigl[\bigl(1-2\theta(\tau)\bigr)\varphi\bigr].  \label{RCpath2}
\end{eqnarray}
This interpolation is a generalization of the interpolation obtained in
Refs.~\cite{RolandCerf,QAB,AQC-intrinsic} by using a local adiabatic condition.


\subsubsection{Constant-gap interpolation}

\label{c-gap}

Rather than assuming condition (\ref{BC-3}), here we consider the case in
which the gap is constant, e.g., $\Delta (\tau )=D(\tau )/J=1$. Hence
combining the following relation [Eq.~(\ref{gap})]: 
\begin{equation}
\Bigl[x_{1}\bigl(\theta (\tau )\bigr)-x_{2}\bigl(\theta (\tau )%
\bigr)\Bigr]^{2}+4rx_{1}\bigl(\theta (\tau )\bigr)x_{2}\bigl(\theta (\tau )%
\bigr)=1,
\end{equation}%
and Eq.~(\ref{relat}) yields 
\begin{eqnarray}
x_{1}\bigl(\theta (\tau )\bigr) &=&\frac{1}{2\sqrt{r(1-r)}}\sin \bigl[2\bigl(%
1-\theta (\tau )\bigr)\varphi \bigr],  \label{constgapx1} \\
x_{2}\bigl(\theta (\tau )\bigr) &=&\frac{1}{2\sqrt{r(1-r)}}\sin \bigl[%
2\theta (\tau )\varphi \bigr].  \label{constgapx2}
\end{eqnarray}


\subsection{Error estimation for different interpolations}

\label{ErPaths}

Having given a general recipe for adiabatic interpolations and having
obtained three particular interpolations, we proceed to compute the
adiabatic error for each of these interpolations. Our main interest here is
to analyze how the run time scales with system size for each of these three
interpolations. We shall also contrast the result for each case with the
estimate obtained from the traditional adiabatic theorem. As a result, we
will demonstrate that the traditional adiabatic condition is not always
reliable for estimation of the minimum run time (given an error threshold),
whereas the time we obtain from the exponential regime of the adiabatic
error is in fact accurate. A remarkable feature of this result is that the
estimated times (in an appropriate sense) need not be very large.


\subsubsection{Linear interpolation}

\begin{figure}[bp]
  \includegraphics[width=8.5cm]{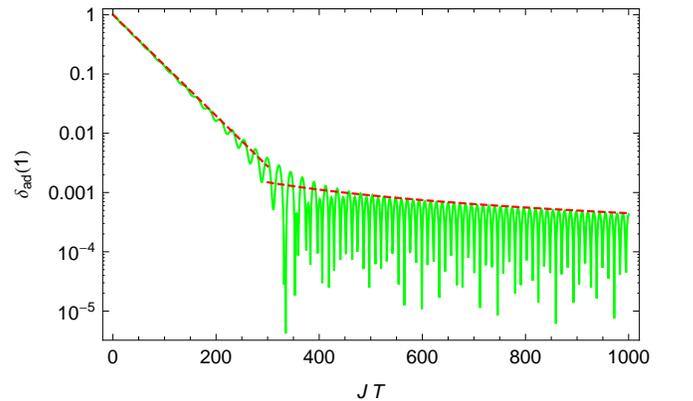}
\caption{(Color online) $\protect\delta _{\text{ad}}(1)$ for the linear
interpolation $\mathbf{x}(\protect\tau )=(1-\protect\tau ,\protect\tau )$,
obtained by numerically solving the corresponding Schr\"{o}dinger equation,
for $r=0.05$. The dashed lines represent the exponential fits $e^{-\protect%
\pi JTr/8}$ [Eq.~(\protect\ref{b2})] and the polynomial fit $2\protect\sqrt{r%
}/(JT)$ [Eq.~(\protect\ref{b1})], from left to right, respectively.}
\label{fig.linear}
\end{figure}
In this case, $\bigl(x_{1}(\tau ),x_{2}(\tau )\bigr)=(1-\tau ,\tau )$, so
that we have 
\begin{eqnarray}
\Vert H(\tau )\Vert /J &\overset{\text{(\ref{Hnorm})}}{=}&1, \\
\Vert \dot{H}(\tau )\Vert /J &\overset{\text{(\ref{Hdotnorm})}}{=}&1, \\
\Delta _{\min } &\overset{\text{(\ref{gap})}}{=}&\sqrt{r}.
\end{eqnarray}%
Hence the traditional adiabatic condition (\ref{cond}) implies that, for $%
\delta _{\text{ad}}(1)\leq \varepsilon $, we should have 
\begin{equation}
T\gg \frac{1}{J\varepsilon r},
\end{equation}%
or equivalently from Eq.~(\ref{runT}), 
\begin{equation}
\tau _{\text{run}}=O\Bigl(\frac{1}{\varepsilon r}\Bigr).
\end{equation}%
That is, that the adiabatic quantum search with a linear interpolation
Hamiltonian requires a run time $O(N)$ [recall $r=M/N$], hence performing no
better than a classical search algorithm \cite{RolandCerf}.

Note, however, that if we truncate the $\mathrm{poly}(1/T)$ expansion (\ref%
{er-simp}) after the first term, in the $r\ll 1$ limit we obtain 
\begin{equation}
T\gtrapprox \frac{2\sqrt{r}}{J\varepsilon },
\label{b1}
\end{equation}%
i.e., $\tau _{\text{run}}=O(\sqrt{N})$, which is not the right estimate.
This illustrates the caveat we discussed in Sec.~\ref{AdError}.

Now we employ the results we developed in Sec.~\ref{AdError} for estimating
the exponential regime of $\delta _{\text{ad}}(1)$ [Eq.~(\ref{xx})]. It is
obvious that for this case there exist no point at which both $x_{1}(\tau
)=1-\tau $ and $x_{2}=\tau $ vanish. Thus, the pole $z_{0}$ (where the gap
vanishes) is obtained simply by inverting Eq.~(\ref{poles}), i.e., 
\begin{equation}
{z_{0}}_{\pm }=\frac{1}{2}\pm \frac{i\sqrt{r}}{2\sqrt{1-r}}.
\end{equation}%
The integral over the gap can be evaluated explicitly as 
\begin{equation}
\mathrm{Im}\left[ \int_{0}^{{z_{0}}_{+}}\Delta (z)~\mathrm{d}z\right] =\frac{%
\pi r}{8\sqrt{1-r}}.
\end{equation}%
For the integral (\ref{residue}), we choose the contour $\mathcal{C}$ to be
a rectangle composed of: (i) $\mathcal{C}_{1}$ the real line $[0,1]$, (ii) $%
\mathcal{C}_{2}$, the line connecting $z=1$ to $z=1+i\infty $, (iii) $%
\mathcal{C}_{3}$, the line connecting $z=1+i\infty $ to $z=i\infty $, and
(iv) $\mathcal{C}_{4}$, the line connecting $z=1+i\infty $ to $z=0$. From
the form of $\widehat{h}_{A}$, we can easily see that $\int_{\mathcal{C}%
_{3}}=0$ [because $\lim_{z\rightarrow \infty }\Delta (z)\rightarrow \infty $%
] and $\int_{\mathcal{C}_{2}}=-\int_{\mathcal{C}_{4}}$ [because $\Delta
(z)=\Delta (1-z)$]. This means that for $r\ll 1$ and in the regime in which $%
|\mathcal{I}_{1}(1)|$ gives the dominant contribution to $\delta _{\text{ad}%
}(1)$, the error exhibits an exponentially decreasing behavior as 
\begin{equation}
\delta _{\text{ad}}(1)\overset{\text{(\ref{residue})}}{\lessapprox
}\frac{\pi }{2}e^{-\pi JTr/8}.
\label{b2}
\end{equation}%
In this regime, for $\delta _{\text{ad}}(1)\leq \varepsilon $ it is
sufficient to have 
\begin{equation}
T\gtrapprox \frac{8\log (1/\varepsilon )}{\pi Jr}+\frac{8\log (\pi/2)}{\pi Jr},
\end{equation}%
whereby we can estimate the following run time: 
\begin{equation}
\tau _{\text{run}}=O\Bigl(\frac{1}{r}\log (1/\varepsilon )\Bigr).
\end{equation}%
This result agrees perfectly with the expected $O(N)$ scaling, with a 
\textit{logarithmic} (rather than inverse) dependence on the error $%
\varepsilon $. Figure~\ref{fig.linear} depicts the adiabatic error vs time,
calculated by solving the corresponding Schr\"{o}dinger equation
numerically. It illustrates the exponential and polynomial regimes.

\textit{Remark.} Here we could not use Eq.~(\ref{st-ph}) because $\widehat{%
h}_{A}$ has poles ${z_{0}}_{\pm }$.


\subsubsection{Constant-norm interpolation}

Here for simplicity, we assume $\theta (\tau )=\tau $. From Eqs.~(\ref%
{RCpath1}) and (\ref{RCpath2}), the gap (\ref{gap}) is 
\begin{equation}
\Delta (\tau )=\sqrt{r}\sec [(1-2\tau )\varphi ],  \label{g-c-n}
\end{equation}%
with the minimum value $\min_{\tau }\Delta =\sqrt{r}$ at $\tau _{\min }=1/2$%
, and Eqs.~(\ref{Hnorm}) and (\ref{Hdotnorm}) yield 
\begin{eqnarray}
&\Vert H(\tau )\Vert /J=&1, \\
&\Vert \dot{H}(\tau )\Vert /J=&\sqrt{r}\varphi \sec ^{2}[(1-2\tau )\varphi ],
\end{eqnarray}%
the latter with the maximum value $\max_{\tau }\Vert \dot{H}(\tau )\Vert
=\varphi /\sqrt{r}$ at $\tau _{\max }=0,1$. Thus, according to the
traditional adiabatic condition (\ref{cond}), in order to have $\delta _{%
\text{ad}}(1)\leq \varepsilon $, we should require 
\begin{equation}
T\gg \frac{1}{J\varepsilon r\sqrt{r}},
\end{equation}%
and in turn, 
\begin{equation}
\tau _{\text{run}}=O\Bigl(\frac{1}{\varepsilon r\sqrt{r}}\Bigr),
\end{equation}%
which is larger than the expected $O(\sqrt{N/M})$ Grover-like scaling \cite%
{RolandCerf,QAB}.
\begin{figure}[bp]
\includegraphics[width=8.5cm]{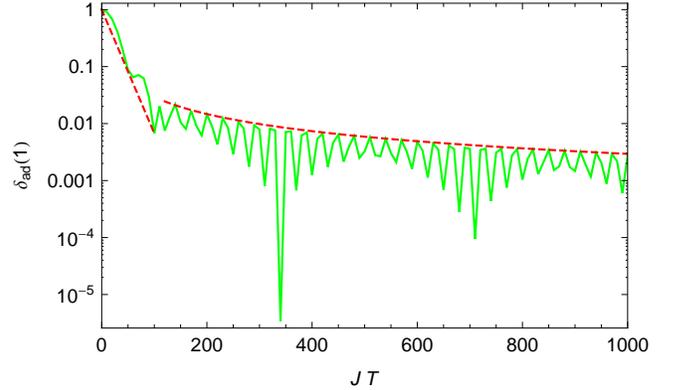}
\caption{(Color online) $\protect\delta _{\text{ad}}(1)$ for the
constant-norm interpolation [Eqs.~(\protect\ref{RCpath1}) and (\protect\ref%
{RCpath2})], obtained by numerically solving the corresponding Schr\"{o}%
dinger equation, for $r=0.01$. The dashed lines represent the exponential
fit $e^{-JT\protect\sqrt{r}/2}$ [Eq.~(\protect\ref{DelLin})] and the
polynomial fit $2\protect\varphi (r)/(JT)$ [Eq.~(\protect\ref{c1})].}
\label{fig.rc}
\end{figure}

On the other hand, truncation of the corresponding $\mathrm{poly}(1/T)$
expansion (\ref{er-simp}) results in 
\begin{equation}
T\gtrapprox \frac{2\varphi }{J\varepsilon },
\label{c1}
\end{equation}%
in which we used $|\dot{x}_{1}(1)|=|\dot{x}_{2}(0)|=\varphi /\sqrt{r(1-r)}$
[Eqs.~(\ref{RCpath1}) and (\ref{RCpath2})] and $\varphi \approx \pi /2$.
Hence, 
\begin{equation}
\tau _{\text{run}}=O(1),
\end{equation}%
which of course is incorrect.

Now we show that a careful treatment of $\mathcal{I}_{1}(\tau )$, as in Sec.~%
\ref{AdError}, results in an exponential adiabatic error, and gives the
correct scaling for the run time. Here, we note that $\widehat{h}_{A}(\tau
)=\varphi $ [Eq.~(\ref{h_A})], from which 
\begin{eqnarray}
&&\hskip-3mm\mathcal{I}_{1}(\tau )=\varphi \int_{0}^{\tau }e^{-iJT\int^{\tau
^{\prime }}\Delta }~\mathrm{d}\tau ^{\prime }=\varphi e^{-i\frac{JT\sqrt{r}}{%
\varphi }~\mathrm{arctanh}\sqrt{1-r}}  \notag  \label{error1} \\
&&\times \int_{0}^{\tau }\left[ \frac{1-\sin (\varphi )+\cos (\varphi )\tan
(\varphi \tau ^{\prime })}{1+\sin (\varphi )-\cos (\varphi )\tan (\varphi
\tau ^{\prime })}\right] ^{-i\frac{JT\sqrt{r}}{2\varphi }}\mathrm{d}\tau
^{\prime },  \notag \\
&&
\end{eqnarray}%
where we used the identity $\mathrm{arctanh}(x)=\frac{1}{2}\ln \bigl|\frac{%
1+x}{1-x}\bigr|$. For $r\ll 1$ [$\sin (\varphi )\approx 1$], this gives rise
to 
\begin{eqnarray}
\hskip-4mm|\mathcal{I}_{1}(\tau )| &\lessapprox &\varphi \left\vert
\int_{0}^{\tau }[\tan (\varphi \tau ^{\prime })]^{-i\frac{JT\sqrt{r}}{%
2\varphi }}\mathrm{d}\tau ^{\prime }\right\vert   \notag
\label{integralbeta} \\
&\approx &\frac{\pi }{2}\int_{0}^{1}[\tan (\pi \tau ^{\prime }/2)]^{-i\frac{%
JT\sqrt{r}}{2\varphi }}\mathrm{d}\tau ^{\prime }+O(\sqrt{r}).  \notag \\
&&
\end{eqnarray}%
Further simplification can be obtained by using the identity $%
\int_{0}^{1}[\tan (\pi \tau ^{\prime })]^{-i\alpha }\mathrm{d}\tau ^{\prime
}=\mathrm{sech}(\pi \alpha /2)$ (for $\alpha \geq 0$) and the inequality $%
\mathrm{sech}(y)\leq 2e^{-y}$ (for $y\geq 0$); hence, 
\begin{equation}
\delta _{\text{ad}}(1)\lessapprox \pi e^{-JT\sqrt{r}/2}+O(\sqrt{r}).
\label{DelLin} 
\end{equation}%
This is the corresponding \emph{exponential} behavior for the constant-norm
Hamiltonian interpolation. It implies that for $\delta _{\text{ad}}(1)\leq
\varepsilon $ it is sufficient to have 
\begin{equation}
T\gtrapprox \frac{2\log (1/\varepsilon )}{J\sqrt{r}},
\end{equation}%
or equivalently, 
\begin{equation}
\tau _{\text{run}}=O\Bigl(\frac{\log (1/\varepsilon )}{\sqrt{r}}\Bigr),
\label{mmm}
\end{equation}%
which is the expected Grover-like $O(\sqrt{N/M})$ scaling \cite%
{RolandCerf,QAB}, but with a logarithmic dependence on the error.

Although Eq.~(\ref{residue}) may not be applicable to the case of
constant-norm interpolation (for $\widehat{h}_{A}$ does not have any
singularity), we can apply Eq.~(\ref{st-ph}) instead. In fact, Eq.~(\ref%
{g-c-n}) implies that 
\begin{equation}
{z_{0}}_{\pm }=\pm i\infty .
\end{equation}%
Hence, 
\begin{equation}
\mathrm{Im}\left[ \int_{0}^{{z_{0}}_{+}}\Delta (z)~\mathrm{d}z\right] =\frac{%
\pi \sqrt{r}}{4\varphi }\overset{r\ll 1}{\approx }\frac{\sqrt{r}}{2},
\end{equation}%
and in turn [Eq.~(\ref{st-ph})] 
\begin{equation}
\delta _{\text{ad}}(1)\lessapprox R(1)~e^{-JT\sqrt{r}/2},
\end{equation}%
with some non-exponentially decreasing $R(1)$. This results in a scaling
similar to Eq.~(\ref{mmm}).


\subsubsection{Constant-gap interpolation}

In this case [$\Delta (\tau )=1$], from Eqs.~(\ref{Hnorm}), (\ref{Hdotnorm}%
), (\ref{constgapx1}), and (\ref{constgapx2}) we obtain 
\begin{eqnarray}
&\Vert H(\tau )\Vert /J=&|\cos [(1-2\tau )\varphi ]|/\sqrt{r}, \\
&\Vert \dot{H}(\tau )\Vert /J=&\varphi |\sin [(1-2\tau )\varphi ]|/\sqrt{r}.
\label{temp1}
\end{eqnarray}%
Hence, according to the traditional adiabatic condition (\ref{cond}), in
order to have $\delta _{\text{ad}}(1)\leq \varepsilon $, it is sufficient to
have 
\begin{equation}
T\gg \frac{\varphi }{J\varepsilon \sqrt{r}},
\end{equation}%
whereby 
\begin{equation}
\tau _{\text{run}}=O\Bigl(\frac{1}{\varepsilon r}\Bigr),  \label{rr1}
\end{equation}%
which is quadratically larger than the Grover-like $O(\sqrt{N/M})$ scaling.

\begin{figure}[bp]
\includegraphics[width=8.5cm]{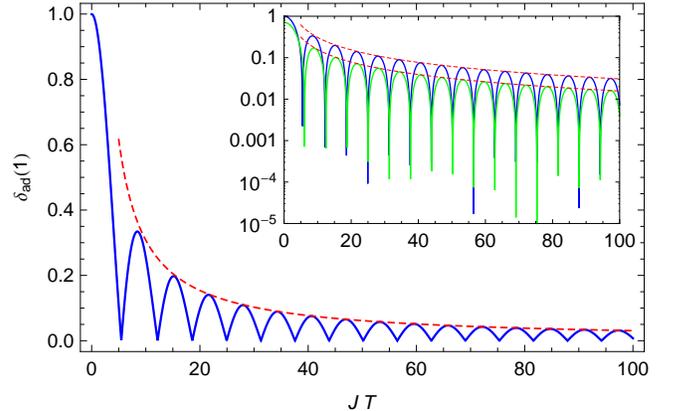}
\caption{(Color online) $\protect\delta _{\text{ad}}(1)$ for the
constant-gap interpolation [Eqs.~(\protect\ref{constgapx1}) and (\protect\ref%
{constgapx2})]. Here, $r=0.001,0.5$ (blue and green, respectively), and the
dashed lines (red) show the $2\protect\varphi (r)/JT$ envelopes. The points $%
T_{k}$, where 
$\protect\delta _{\text{ad}}(1)\bigl|_{T_{k}}=0$, 
are given by Eq.~(\protect\ref{T_k}).}
\label{fig.constg}
\end{figure}

On the other hand, noting that $|\dot{x}_{1}(1)|=|\dot{x}_{2}(0)|=\varphi /%
\sqrt{r(1-r)}$ [Eqs.~(\ref{constgapx1}) and (\ref{constgapx2})], the
truncation of the corresponding $\mathrm{poly}(1/T)$ expansion after the
first term yields 
\begin{equation}
T\gtrapprox \frac{2\varphi }{J\varepsilon },  \label{temp4}
\end{equation}%
or equivalently 
\begin{equation}
\tau _{\text{run}}=O\Big(\frac{1}{\varepsilon \sqrt{r}}\Bigr).  \label{rr2}
\end{equation}%
Clearly, Eqs.~(\ref{rr1}) and (\ref{rr2}) are not in agreement.

It is interesting to note that here one can in fact solve the Schr\"{o}%
dinger equation exactly. The operator $\widehat{\Omega }_{T}(\tau )$ [Eq.~(%
\ref{o-hat})] satisfies the following equation: 
\begin{equation}
\dot{\widehat{\Omega }}=-i\widehat{W}\widehat{\Omega },  \label{temp3}
\end{equation}%
in which 
\begin{eqnarray}
\widehat{W} &=&T\widehat{A}^{\dag }\widehat{H}\widehat{A}-\widehat{A}^{\dag }%
\widehat{H}_{A}\widehat{A}  \notag \\
&\overset{\text{(\ref{constgapx1}),(\ref{constgapx2}),(\ref{h_A})}}{=}&\frac{%
JT}{2}\left( \frac{\cos [(1-2\tau )\varphi ]}{\sqrt{r}}\openone+\sigma
_{z}\right) -\varphi \sigma _{y}.  \notag \\
&&
\end{eqnarray}%
Since $\widehat{W}$ is time-independent, integration of Eq.~(\ref{temp3}) is
straightforward: 
\begin{eqnarray}
\widehat{\Omega }_{T}(\tau ) &=&e^{-i\int_{0}^{\tau }\widehat{W}(1)\mathrm{d}%
\tau ^{\prime }}=e^{-\frac{i\tau JT\sqrt{1-r}}{2\varphi \sqrt{r}}}  \notag
\label{constgapsol} \\
&&\times \frac{\sin \bigl(\tau \sqrt{\varphi ^{2}+(JT)^{2}/4}\bigr)}{\sqrt{%
\varphi ^{2}+(JT)^{2}/4}}\Bigl[\sqrt{\varphi ^{2}+(JT)^{2}/4}  \notag \\
&&\times \cot \bigl(\tau \sqrt{\varphi ^{2}+(JT)^{2}/4}\bigr)\openone%
+i\varphi \tau \sigma _{y}-\frac{iJT}{2}\sigma _{z}\Bigr].  \notag \\
&&
\end{eqnarray}%
Thus, from Eq.~(\ref{ad-error}) the adiabatic error is exactly
\begin{equation}
\hskip-3mm\delta _{\text{ad}}(1)=\varphi \frac{\bigl|\sin \sqrt{\varphi
^{2}+(JT)^{2}/4}\bigr|}{\sqrt{\varphi ^{2}+(JT)^{2}/4}}.
\label{c-g-T}
\end{equation}

Figure~\ref{fig.constg} depicts $\delta _{\text{ad}}(1)$ for two different
values of $r$. Note that for large evolution times $T\gg 2\varphi /J$, we
obtain 
\begin{equation}
\delta _{\text{ad}}(1)\leq \frac{2\varphi }{JT}.
\label{cg-1}
\end{equation}%
This implies that, in the $r\ll 1$ limit, in order for $\delta _{\text{ad}%
}(1)\leq \varepsilon $ it is sufficient to have 
\begin{equation}
\tau _{\text{run}}=O\Bigl(\frac{1}{\varepsilon \sqrt{r}}\Bigr),
\end{equation}%
which is the Grover-like scaling $O(\sqrt{N/M})$. Notice that in this limit
the adiabatic error behaves inverse-linearly, $\delta _{\text{ad}}(1)\leq
2\varphi /JT$, which is in perfect agreement with Eq.~(\ref{temp4}). In
addition, we observe that there exist $T$s less than the above limit in in
which the adiabatic error can vanish (hence instantaneous full
adiabaticity). According to Eq.~(\ref{c-g-T}), we have 
$\delta _{\text{ad}}(1)\bigl|_{T_{k}}=0$ 
where \begin{equation}
JT_{k}=2\sqrt{k^{2}\pi ^{2}-\varphi ^{2}}\overset{r\ll 1}{\approx }2\pi 
\sqrt{k^{2}-1/4},
 \label{T_k}
\end{equation}%
for $k\in \mathbb{N}$. The existence of such $T_{k}$s is in agreement with
Ref.~\cite{LZ-Avron}. Figure~\ref{fig.constg} shows $\delta _{\text{ad}}(1)$
for two different values of $r$.

\textit{Remark.} As is evident here the adiabatic error does not show any
exponential behavior. In fact, neither of the methods we discussed in
subsection~\ref{expl} is applicable.


\subsubsection{General interpolation}

\label{g-interp}

Here, we discuss the behavior of the exponential $e^{-JT\mathrm{Im}%
[\int^{z_{0}}\Delta ]}$ for the general interpolation we derived in
subsection~\ref{path-g}. Our analysis is based on a formal power series
expansion of $x_{1}\big(\theta (\tau )\bigl)$ and $x_{2}\big(\theta (\tau )%
\bigl)$ in terms of $r$---recall that we are interested in the regime $r\ll 1
$. We further assume that $\theta (\tau )$ does not depend explicitly on $r$.

Consider the following formal expansions: 
\begin{eqnarray}
\hskip-6mm &&x_{1}\bigl(\theta (\tau )\bigr)=f_{1}\bigl(\theta (\tau )\bigr)%
+g_{1}\bigl(\theta (\tau )\bigr)r^{\alpha _{1}}+O(r^{\alpha _{1}+\epsilon
_{1}}),  \label{e1} \\
\hskip-6mm &&x_{2}\bigl(\theta (\tau )\bigr)=f_{2}\bigl(\theta (\tau )\bigr)%
+g_{2}\bigl(\theta (\tau )\bigr)r^{\alpha _{2}}+O(r^{\alpha _{2}+\epsilon
_{2}}),  \label{e2}
\end{eqnarray}%
in which $\alpha _{1}$ and $\alpha _{2}$ are some nonnegative numbers (to be
determined later), $\epsilon _{1},\epsilon _{2}>0$, and $f_{1},f_{2}\neq 0$.
We notice that the linear and constant-gap interpolations (subsections~\ref%
{lin-sec} and \ref{c-gap}) do not admit expansions as in Eqs.~(\ref{e1}) and
(\ref{e2}). Equation (\ref{gap}) hence yields 
\begin{eqnarray}
\Delta ^{2} &=&(f_{1}-f_{2})^{2}+2(f_{1}-f_{2})(g_{1}r^{\alpha
_{1}}-g_{2}r^{\alpha _{2}})+4f_{1}f_{2}r  \notag \\
&&+(g_{1}r^{\alpha _{1}}-g_{2}r^{\alpha _{2}})^{2}+O(r^{\alpha _{1}+\alpha
_{2}+\epsilon _{1}+\epsilon _{2}}).  \label{delta-r}
\end{eqnarray}%
Similarly, inserting Eqs.~(\ref{e1}) and (\ref{e2}) into $x_{21}=x_{2}/x_{1}$
gives 
\begin{equation}
x_{21}=\frac{f_{2}}{f_{1}}-\frac{f_{2}g_{1}}{f_{1}^{2}}r^{\alpha _{1}}+\frac{%
g_{2}}{f_{1}}r^{\alpha _{2}}+O(r^{\alpha _{1}+\alpha _{2}+\epsilon
_{1}+\epsilon _{2}}).
\label{x21-1}
\end{equation}%
On the other hand, Eq.~(\ref{relat}) yields 
\begin{equation}
x_{21}(\tau )=1-2\sqrt{r}\cot \bigl(\pi \theta (\tau )\bigr)+O(r).
\label{x21-2}
\end{equation}%
The symmetry $x_{2}\bigl(\theta (\tau )\bigr)=x_{1}\bigl(1-\theta (\tau )%
\bigr)$ [Eq.~(\ref{symmetry})] requires that 
\begin{equation}
g_{1}\bigl(\theta (\tau )\bigr)=g_{2}\bigl(1-\theta (\tau )\bigr),
\end{equation}%
which in turn implies $\alpha _{1}=\alpha _{2}=1/2$. Comparing the terms
with the same powers of $r$ in Eqs.~(\ref{x21-1}) and (\ref{x21-2}), we
conclude that 
\begin{equation}
f_{1}\bigl(\theta (\tau )\bigr)=f_{2}\bigl(\theta (\tau )\bigr)\equiv f\bigl(%
\theta (\tau )\bigr),
\end{equation}%
and 
\begin{equation}
\frac{g_{1}\bigl(\theta (\tau )\bigr)-g_{2}\bigl(\theta (\tau )\bigr)}{f%
\bigl(\theta (\tau )\bigr)}=2\cot \bigl(\pi \theta (\tau )\bigr).
\label{fg-eq}
\end{equation}%
After inserting the above relations back into Eq.~(\ref{delta-r}) and using
Eq.~(\ref{fg-eq}), we obtain 
\begin{equation}
\Delta (\tau )=2f\bigl(\theta (\tau )\big)\csc \bigl(\pi \theta (\tau )\bigr)%
\sqrt{r}+O(r).
\label{gg-0}
\end{equation}%
Now we assume that $f\bigl(\theta (\tau )\bigr)\neq 0$ everywhere, or if
there exist points at which $f$ vanishes, 
their contribution to the integral $\mathrm{Im}[\int^{z_{0}}\Delta ]$
is not substantial. Note that the previous condition is in fact a
condition on the norm of the Hamiltonian---because from Eq.~(\ref{Hnorm}) 
\begin{equation}
\Vert H\Vert /J=2f+O(\sqrt{r}).
\label{Hnorm-f}
\end{equation}%
Therefore, provided that for large times the adiabatic error asymptotically
behaves as in Eq.~(\ref{residue}) or (\ref{st-ph}), we obtain 
\begin{equation}
\hskip-1mm\delta _{\text{ad}}(1)\lessapprox R(1)e^{-2\sqrt{r}JT\bigl|\mathrm{%
Im}\bigl[\int_{0}^{z_{0}}f\bigl(\theta (z)\bigr)\csc \bigl(\pi \theta (z)%
\bigr)~\mathrm{d}z\bigr]\bigr|},
\end{equation}%
where $R(1)$ is a non-exponential function of $T$ (which may also depend
weakly on $r$). Hence, in the $r\ll 1$ limit for $\delta _{\text{ad}}(1)\leq
\varepsilon $, it is sufficient to have 
\begin{equation}
T\gtrapprox \frac{\log (1/\varepsilon )}{2J\sqrt{r}}\Bigl|\mathrm{Im}\Bigl[%
\int_{0}^{z_{0}(r)}f\bigl(\theta (z)\bigr)\csc \bigl(\pi \theta (z)\bigr)~%
\mathrm{d}z\Bigr]\Bigr|.
\end{equation}%
This in turn implies the following scaling for the run time: 
\begin{eqnarray}
\tau _{\text{run}} &=&O\Bigl(\frac{\log (1/\varepsilon )}{\sqrt{r}}\Bigl|%
\mathrm{Im}\Bigl[\int_{0}^{z_{0}(r)}f\bigl(\theta (z)\bigr)\csc \bigl(\pi
\theta (z)\bigr)~\mathrm{d}z\Bigr]\Bigr|\Bigr).  \notag  \label{generalT} \\
&&
\end{eqnarray}%
The overall $r$-dependence here comes from $\sqrt{r}$ and $z_{0}(r)$; e.g.,
we recover the Grover-like $O(\sqrt{N/M})$ scaling if $z_{0}$ does not
depend on $r$. This analysis then highlights in a fairly general way the
interplay between $r$, $T$, and $\delta _{\text{ad}}(1)$ in the quantum
search algorithm.


\subsection{A strategy for reducing the adiabatic error}

\label{strategy}

For most applications it is desirable to make the adiabatic error as small
as possible. We have seen that $\delta _{\text{ad}}(1)$ contains exponential
terms, suppressed by the polynomially-decaying terms. Therefore, it is
useful to somehow extend the dominance of the exponential term by reducing
the contribution of the polynomial term, e.g., by prolonging the dominance
of the exponential regime or by enforcing higher order polynomial behavior.
In the following we shall discuss a control strategy for reducing the
adiabatic error by manipulation of the boundary conditions (see, e.g., Refs.~%
\cite{HagedornJoye:02,LRH}).


\subsubsection{A general strategy: Control via boundary conditions}

Equation~(\ref{er-simp}) demonstrates explicitly how the adiabatic error
depends on the boundary conditions, up to $O(1/T)$. Interestingly, if we
choose $\dot{x}_{2}(0)=\dot{x}_{1}(1)=0$, the prefactor of the $1/T$
vanishes, whence $\delta _{\text{ad}}(1)=O(1/T^{2})$. In a similar fashion,
one can see that by enforcing suitable (extra) boundary conditions on the
interpolation $\mathbf{x}(\tau )$ the prefactor of the $1/T^{2}$ or even
higher order terms can be made zero. This implies that by manipulating the
boundary conditions of the interpolation, one may achieve smaller adiabatic
errors. This observation is a manifestation of the following general
theorem: If the Hamiltonian $H(\tau )$ is sufficiently differentiable,
forcing \textit{all} time derivatives of the Hamiltonian up to some order $k$
to vanish at the boundaries,%
\begin{equation}
\frac{\mathrm{d}^{l}H(\tau )}{\mathrm{d}\tau ^{l}}\Bigl|_{\tau \in
\{0,1\}}=0~~~\forall l\in \{1,\ldots ,k\},  \label{HDZ}
\end{equation}%
is sufficient for $\delta _{\text{ad}}(1)=O(1/T^{k+1})$ \cite%
{Sancho,HagedornJoye:02,JRS,LRH}. It is interesting to note that the very
same condition together with the assumption of the \emph{analyticity} of $%
H(\tau )$ in a small strip around the real axis in the \emph{complex} $\tau $%
-plane give rise to $\delta _{\text{ad}}(1)=O(e^{-c(r)JT})$, where $%
c(r)\equiv \Delta _{\min }^{3}/\max_{\tau }\Vert \dot{H}(\tau )\Vert ^{2}$
[up to an $O(1)$ prefactor] \cite{LRH}.\footnote{%
Note that the conditions $\dot{x}_{2}(0)=\dot{x}_{1}(1)=0$ we obtained above
are in fact weaker than requiring $\dot{x}_{1}(\{0,1\})=\dot{x}%
_{2}(\{0,1\})=0$ [sufficient for $\dot{H}(\tau )|_{\tau \in \{0,1\}}=0$].}

This is a remarkable result, in that it guarantees that with sufficient
smoothness at two points one can substantially suppress the adiabatic error.
This is a fairly low price to pay for higher accuracy. In particular, in
experimental realizations, manipulating Hamiltonian interpolations only at
the beginning and the end (as opposed to from beginning to end), may offer a
less demanding control strategy than one seeking to control the dynamics
instantaneously along the entire evolution.

Now we show that in the framework we developed earlier, enforcing the
required smoothness properties can be achieved by choosing 
an appropriate $\theta (\tau )$ function. We recall that this function was fairly arbitrary;
we required that it be a monotonically increasing differentiable function ($%
\dot{\theta}(\tau )>0$) satisfying the boundary conditions $\theta (0)=0$
and $\theta (1)=1$. We require further that $\theta (\tau )\equiv \theta
_{k}(\tau )$ (for a given $k\in \mathbb{N}$) have the following property: 
\begin{equation}
\frac{\mathrm{d}^{l}\theta _{k}(\tau )}{\mathrm{d}\tau ^{l}}\Bigl|_{\tau \in
\{0,1\}}=0~~~\forall l\in \{1,\ldots ,k\},  \label{theta-cond}
\end{equation}%
namely, the first $k$ derivatives should vanish at the boundaries. This
property is sufficient for fulfilling Eq.~(\ref{HDZ}) because $\dot{H}=\dot{%
\theta}\partial _{\theta }H$. An example of such $\theta _{k}(\tau )$ is the
regularized incomplete beta function, 
\begin{equation}
\theta _{k}(\tau )=\frac{\mathrm{B}_{\tau }(1+k,1+k)}{\mathrm{B}_{1}(1+k,1+k)%
},  \label{beta}
\end{equation}%
in which $\mathrm{B}_{\tau }(a,b)\equiv \int_{0}^{\tau }y^{a-1}(1-y)^{b-1}%
\mathrm{d}y$, with $\mathrm{Re}(a),\mathrm{Re}(b)>0$, and $|\tau |\leq 1$ 
\cite{Arfken:book,Hayek:book}.


\subsubsection{Polynomial reduction}

To demonstrate explicitly how the conditions (\ref{HDZ}) affect the
adiabatic error, we employ the method developed in Refs.~\cite%
{HagedornJoye:02,LRH} for approximating the coefficients in the $\mathrm{poly%
}(1/T)$ expansion of $\delta _{\text{ad}}(1)$. One can construct an
approximate (unnormalized) ansatz for the solution to the Schr\"{o}dinger
equation (\ref{eqscalt}) in the powers of $1/T$ as follows: 
\begin{eqnarray}
|\Psi _{n}(\tau )\rangle  &=&e^{-iJT\int^{\tau }E_{0}}\Bigl[|\Phi _{0}(\tau
)\rangle +\sum_{l=1}^{n}\frac{1}{(JT)^{l}}|\psi _{l}(\tau )\rangle   \notag
\\
&&+\frac{|\psi _{n+1}^{\perp }(\tau )\rangle }{(JT)^{n+1}}\Bigr],
\end{eqnarray}%
with the error 
\begin{equation}
\Vert |\psi _{T}(1)\rangle -|\Psi _{n}(1)\rangle \Vert \leq \frac{%
\max_{\tau }\Vert |\dot{\psi}_{n+1}^{\perp }(\tau )\rangle \Vert }{(JT)^{n+1}%
}.
\end{equation}%
Here, $\{|\psi _{l}(\tau )\rangle \}$ and $\{|\psi _{l}^{\perp }(\tau
)\rangle \}$ are given as follows \cite{LRH}: 
\begin{eqnarray}
|\psi _{l}\rangle  &=&a_{l}|\Phi _{0}\rangle +|\psi _{l}^{\perp }\rangle , \\
|\psi _{l}^{\perp }\rangle  &=&G_{r}(f_{l-1}|\dot{\Phi}_{0}\rangle +|\dot{%
\psi}_{l-1}^{\perp }\rangle ), \\
a_{l} &=&-\int_{\tau }^{1}\langle \Phi _{0}|\dot{\psi}_{l}^{\perp }\rangle ~%
\mathrm{d}\tau ^{\prime },~a_{0}\equiv 1,  \label{fs} \\
G_{r} &=&i(H-E_{0})^{-1}(\openone-|\Phi _{0}\rangle \langle \Phi _{0}|).
\end{eqnarray}%
It is evident that $|\psi _{T}(\tau )\rangle =|\Psi _{\infty }(\tau )\rangle
/\Vert \Psi _{\infty }(\tau )\Vert $.

Provided that all $n$ derivatives of $H(\tau )$ vanish at the boundaries,
Eq.~(\ref{HDZ}), all the terms except $|\Phi _{0}(1)\rangle $ and $|\psi
_{n+1}^{\perp }(1)\rangle /(JT)^{n+1}$ will vanish \cite{Sancho} (see also
Ref.~\cite{LRH}, where with a condition different than Eq.~(\ref{fs}) all
the terms except $|\Phi _{0}\rangle $ vanish).

Let us define 
\begin{eqnarray}
&&\hskip-8mm\delta _{1}(1)\equiv \sqrt{1-|\langle \widehat{\psi }_{T}(1)|%
\widehat{\Psi }_{n}(1)\rangle |^{2}/\Vert \widehat{\Psi }_{n}(1)\Vert ^{2}},
\label{delta1-T} \\
&&\hskip-8mm\delta _{2}(1)\equiv \sqrt{1-|\langle \widehat{\Psi }_{n}(1)|%
\widehat{\Phi }_{-}(1)\rangle |^{2}/\Vert \widehat{\Psi }_{n}(1)\Vert ^{2}}.
\label{delta2-T}
\end{eqnarray}%
Since $\delta (a,b)\equiv \sqrt{1-|\langle a|b\rangle |^{2}}$ (for
normalized $|a\rangle $ and $|b\rangle $) is a distance [Eq.~(\ref{delta-D}%
)], from the triangle inequality $\delta (a,b)\leq \delta (a,c)+\delta (b,c)$
we have: 
\begin{equation}
\delta _{\text{ad}}(1)\leq \delta _{1}(1)+\delta _{2}(1).
\end{equation}%
%
%
%
%
Notice that $\delta _{1}(1)$ can also be written as follows: 
\begin{eqnarray}
\delta _{1}(1) &=&\sqrt{1-\frac{|\langle \widehat{\Psi }_{n}(1)|\widehat{%
\Psi }_{\infty }(1)\rangle |^{2}}{\Vert \widehat{\Psi }_{n}(1)\Vert
^{2}\Vert \widehat{\Psi }_{\infty }(1)\Vert ^{2}}}  \notag \\
&=&\sqrt{1-\frac{|\langle \widehat{\Psi }_{n}(1)|(|\widehat{\Psi }%
_{n}(1)\rangle +|\widehat{R}_{n}(1)\rangle )|^{2}}{%
[1+O(1/T)]^{2}[1+O(1/T)]^{2}}}  \notag \\
&\approx &\sqrt{1-(1+2\mathrm{Re}[\langle \widehat{\Psi }_{n}(1)|\widehat{R}%
_{n}(1)\rangle ])}  \notag \\
&=&O\Bigl(\frac{1}{T^{n+2}}\Bigr),  \label{delta1-T-2}
\end{eqnarray}%
where 
\begin{equation}
|\widehat{R}_{n}(1)\rangle \equiv |\widehat{\Psi }_{\infty }(1)\rangle -|%
\widehat{\Psi }_{n}(1)\rangle .
\end{equation}

\begin{figure}[bp]
\includegraphics[width=8.5cm]{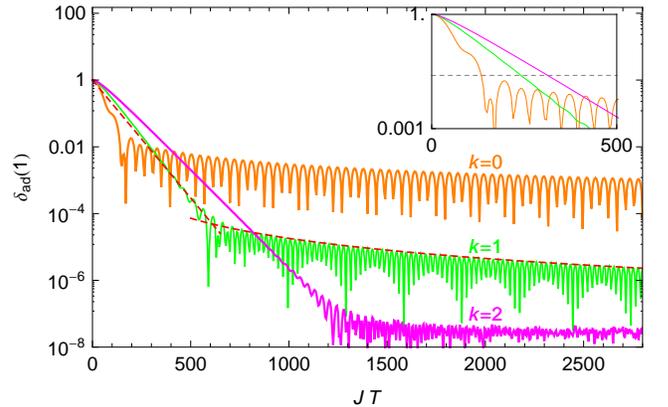}
\caption{(Color online) $\protect\delta _{\text{ad}}(1)$ obtained by
numerically solving the Schr\"{o}dinger equation corresponding to the
constant-norm interpolation with $\protect\theta _{k}(\protect\tau )$ the
regularized beta function [Eq.~(\protect\ref{beta})], for $k\in \{0,1,2\}$
(here $r=2^{-8}$). The dashed (red) lines represent $e^{-\protect\pi JT%
\protect\sqrt{r}/(8\protect\varphi )}$ and $2\protect\varphi |\protect\theta %
_{1}^{(2)}(1)|/(JT)^{2}$ [Eq.~(\protect\ref{ER})].}
\label{fig:ErrorBeta}
\end{figure}

On the other hand, a straightforward calculation (supplemented with
induction) shows that 
\begin{eqnarray}
\delta _{2}(1) &=&\sqrt{1-|\langle \widehat{\Psi }_{n}(1)|\widehat{\Phi }%
_{-}(1)\rangle |^{2}/\Vert \widehat{\Psi }_{n}(1)\Vert ^{2}}  \notag \\
&=&\frac{\sqrt{r(1-r)}|1-2r|}{(JT)^{n+1}}\Bigl(\bigl|\dot{x}_{2}(0)\theta
_{n}^{(n+1)}(0)\bigr|  \notag \\
&&+\bigl|\dot{x}_{1}(1)\theta _{n}^{(n+1)}(1)\bigr|\Bigr)+O\Bigl(\frac{1}{%
T^{n+2}}\Bigr).  \label{delta2}
\end{eqnarray}%
This bound holds true for any interpolating paths $\mathbf{x}(\tau )$ for
which $x_{1}(1-\tau )=x_{2}(\tau )$. As can be seen, $\delta _{1}(1)$ is
negligible in comparison to $\delta _{2}(1)$, hence we obtain 
\begin{eqnarray}
\delta _{\text{ad}}(1) &\leq &\frac{\sqrt{r(1-r)}|1-2r|}{(JT)^{n+1}}\Bigl(%
\bigl|\dot{x}_{1}(1)\theta _{n}^{(n+1)}(1)\bigr|  \notag \\
&&+\bigl|\dot{x}_{2}(0)\theta _{n}^{(n+1)}(0)\bigr|\Bigr)+O\Bigl(\frac{1}{%
T^{n+2}}\Bigr).
\end{eqnarray}%
This result is a generalization of Eq.~(\ref{er-simp}). For example, in the
case of the constant-gap interpolation (subsection~\ref{c-gap}), this error
reduces to 
\begin{equation}
\delta _{\text{ad}}(1)\lessapprox \frac{2\varphi \bigl|\theta _{n}^{(n+1)}(1)%
\bigr|}{(JT)^{n+1}},  \label{ER}
\end{equation}%
in comparison with Eq.~(\ref{cg-1}). Figure~\ref{fig:ErrorBeta} depicts $%
\delta _{\text{ad}}(1)$ for the constant-norm interpolation [Eqs.~(\ref%
{RCpath1}) and (\ref{RCpath2})]. It can be seen that by increasing $k$ the
exponential regime dominates longer, while the polynomial regime is pushed
farther away to the region of large evolution times. However, this
improvement comes at a price. The rate of exponentiality decreases with
increasing $k$; that is, if $k_{1}<k_{2}$ and $\delta _{\text{ad}}(1)\leq
\varepsilon $ for an $\varepsilon $ in the exponential regime for the larger 
$k$, then $T_{1}(\varepsilon )>T_{2}(\varepsilon )$ [see the inset of Fig.~%
\ref{fig:ErrorBeta}]. In other words, for some values of $\varepsilon $,
increasing $k$ might give rise to an increased run time. Of course, if $%
\varepsilon $ is such that the polynomial regimes dominate for both values
of $k_{1}$ and $k_{2}$, the interpolation with the larger $k$ ($k_{2}$)
results in a smaller run time. 


\subsubsection{Exponential reduction}

Since choosing a $\theta _{k}(\tau )$ with a larger $k$ benefits the
accuracy of the adiabatic evolution, it is natural to investigate cases with 
$k=\infty $. An example of such $\theta _{\infty }(\tau )$ is 
\begin{equation}
\theta _{\infty }(\tau )=\frac{\int_{0}^{\tau }b_{\alpha \beta }(\tau
^{\prime })~\mathrm{d}\tau ^{\prime }}{\int_{0}^{1}b_{\alpha \beta }(\tau
^{\prime })~\mathrm{d}\tau ^{\prime }},  \label{sing}
\end{equation}%
where 
\begin{equation}
b_{\alpha \beta }(\tau )=e^{-\beta /[\tau ^{\alpha }(1-\tau )^{\alpha
}]}~~~~0<\alpha ,\beta \leq 1,
\end{equation}%
is a symmetric ``bump function". 
We notice that $b_{\alpha \beta }(\tau )$ is compactly-supported and
infinitely differentiable in $\tau \in \lbrack 0,1]$ (the ``Schwartz 
class'' \cite{Gelfand:book}); in particular, 
\begin{equation}
\frac{\mathrm{d}^{l}b_{\alpha \beta }(\tau )}{\mathrm{d}\tau ^{l}}\Bigl|%
_{\tau \in \{0,1\}}=0~~~\forall l\in \mathbb{N}.
\end{equation}%
However, it is not an \emph{analytic} function of $\tau $, which implies
that $H\left( \mathbf{x}(\theta _{\infty }(\tau ))\right) $ is not analytic
either. Despite this infinite smoothness, the very lack of analyticity in
fact prevents the adiabatic error from being identically zero \cite%
{HagedornJoye:02,JRS}.

Nevertheless, $\theta _{\infty }(\tau )$ helps remove the polynomial terms
arbitrarily, hence extending the exponential regime farther. Additionally,
the exponent of the exponential term is controllable through varying the
parameters $\alpha $ and $\beta $. In this case, asymptotic evaluation of
the integral (\ref{IntTran}) with the stationary phase method results in a
faster-than-polynomial convergence to zero. For example, with $\alpha =1$
and $\beta \ll 1$ for the constant-norm interpolation, we can approach an
exponential convergence similarly to the case with $k=0$. The achievement of
an exponentially small $\delta _{\text{ad}}(1)$ for such functions 
may be understood as an instance of rigorously derived exponentially small errors for a general
class of functions called the Gevrey class \cite{Nenciu:93,Martinez}.

It is evident that manipulating the $\theta (\tau )$ function may also
affect the exponent $\mathrm{Im}[\int_{0}^{z_{0}}\Delta \bigl(\theta (\tau
^{\prime })\bigr)\mathrm{d}\tau ^{\prime }]$. This can be observed in Fig.~%
\ref{fig:ErrorBeta} (the inset) through the change of the slope of the
exponential lines in the log plot. A side consequence of using a $\theta
(\tau )$ with the desired boundary conditions (\ref{theta-cond}) is that
increasing $k$ may adversely increase the value of $T$ for which $\delta _{%
\text{ad}}(1)\leq \varepsilon $ (in the exponential regime). To quantify how
choosing a $\theta _{k}(\tau )$ for the Hamiltonian interpolation affects
the performance of the algorithm, we propose the following measure: 
\begin{equation}
\eta _{k}(1)\equiv \frac{1}{T}\int_{0}^{T}\delta _{\text{ad}}(1)\bigl|_{%
\mathbf{x}\bigl(\theta _{k}(\tau )\bigr)}~\mathrm{d}T,  \label{measure}
\end{equation}%
namely, the average adiabatic error up to time $T$, for a given $k$. A
larger average error may be interpreted as less efficient performance.

The above problem with the effect of $k$ on exponentiality may be partially
alleviated in some cases. Recall that in subsection~\ref{path-g} we found a
fairly general interpolation, which resulted in the parametrization $%
x_{12}(\tau )$ [Eq.~(\ref{relat})]---it was later that we added further
conditions so as to find $x_{1}(\tau )$ and $x_{2}(\tau )$ separately.
Rather than assuming the condition~(\ref{BC-3}), let us impose 
\begin{equation}
2f\bigl(\theta (\tau )\bigr)=1+\zeta \dot{\theta}(\tau ),
\end{equation}%
for some $\zeta \neq 0$. We should be mindful of the fact that, from Eq.~(%
\ref{Hnorm-f}), modifying $f$ leads to a modification of the norm of the
Hamiltonian, or equivalently the maximum energy of the system; this is a
cost, which should be taken care of in the correct estimation of $\tau _{%
\text{run}}$ [Eq.~(\ref{runT})]. Additionally, from Eq.~(\ref{gg-0}) it is
seen that---since we assumed $f\neq 0$---(up to the leading order in $r$)
the gap $\Delta $ vanishes where $\sin \bigl(\theta (z_{0})\bigr)\propto
\infty $; i.e., $\theta ({z_{0}}_{\pm })=\pm i\infty $. In this case, the
exponent $\mathrm{Im}[\int^{z_{0}}\Delta \bigl(\theta (\tau ^{\prime })\bigr)%
~\mathrm{d}\tau ^{\prime }]$ [Eq.~(\ref{xx})] becomes 
\begin{eqnarray}
\mathrm{Im}\left[\int_0^{{z_0}_{+}}\bigl(1+\zeta\dot{\theta}(\tau')\bigr)\csc\bigl(\pi\theta\bigr(\tau'))~\mathrm{d}\tau'\right]
=\nonumber\\ 
\mathrm{Im}\left[\int_0^{{z_0}_+}\csc\bigl(\pi\theta(\tau')\bigr)~\mathrm{d}\tau'\right]
+\frac{\zeta}{2}.
\label{mm}
\end{eqnarray}%
Therefore, by appropriately choosing $\zeta $---subject to the condition $%
1+\zeta \dot{\theta}>0$---we can tune the exponent of the exponential term
in $\delta _{\text{ad}}(1)$. This in turn gives us control over the run time
in the exponential regime. Furthermore, as we argued earlier, replacing $%
\theta (\tau )\rightarrow \theta _{k}(\tau )$ causes the polynomial terms of 
$\delta _{\text{ad}}(1)$ to be $O(1/T^{k+1})$. Thus, we are now in
possession of two control parameters $k$ [more precisely $\theta _{k}(\tau )$%
] and $\zeta $ with which we can manipulate how the adiabatic error behaves
in either of the exponential and polynomial regimes. This type of control
may have applications in experiments in which adiabaticity plays a role.

As an example, let $\theta _{k}(\tau )$ be the regularized beta function
[Eq.~(\ref{beta})]. For $k\gg 1$ and after employing Stirling's
approximation for the factorial function ($k!\approx \sqrt{2\pi k}(k/e)^{k}$ 
\cite{Arfken:book}), we obtain 
\begin{eqnarray}
\max_{\tau }\dot{\theta}(\tau ) &=&\dot{\theta}(1/2)\approx \frac{2(2+k)^{-%
\frac{5}{2}-k}(\frac{5}{2}+k)^{3+k}}{\sqrt{\pi e}}\approx \sqrt{k},  \notag
\\
&&
\end{eqnarray}%
whereby 
\begin{equation}
\max_{\tau }\Vert H(\tau )\Vert /J\approx 1+\sqrt{k}|\zeta _{k}|.
\end{equation}%
As a result, for example, in order to keep the maximum energy constant,
while having the advantages of $\theta _{k}$, we should choose $|\zeta
_{k}|=O(1/\sqrt{k})$.


\section{Summary and conclusions}

\label{Summ}

Adiabatic evolution is characterized by a tradeoff between the total time
taken and the error in the final state reached, relative to the desired
adiabatic state. Motivated by a desire to understand and optimize this
tradeoff, in this work we performed a detailed analysis of the adiabatic
error for the case of an adiabatic quantum search algorithm. Rather than
using the traditional adiabatic condition, with its associated pitfalls, we
chose to calculate the adiabatic error directly by solving the Schr\"{o}%
dinger equation. This enabled us to derive an exact relation for the
adiabatic error. Building on this exact result, we employed a formal
polynomial series expansion in $1/T$ for calculating the error term by term.
This also allowed us to bound the adiabatic error. We showed that the
polynomial expansion should be truncated carefully if we aim to obtain a
reliable estimate for the run time of the algorithm.

We demonstrated that employing a different technique based on complex
analysis reveals, in fairly general situations, a regime of exponential
time-dependent decay of the adiabatic error, preceding a polynomial regime.
The latter has been shown to be a general feature of adiabatic Hamiltonians,
whereas the existence of an exponential precursor is not always guaranteed. We
showed how, in case these two regimes both exist, one can estimate the
crossover region---i.e., the time at which a transition between the two
regimes takes place. Equipped with this, we provided an estimate for
the minimum time required for the algorithm to achieve a given
accuracy threshold. Discerning the exponential regime enabled us to give an improved
total time estimate, circumventing the overestimate arising from the error
bound using only the polynomial expansion. Indeed, the total time estimated
from the exponential regime always gave the correct scaling with system size
(the well-known quadratic speedup over classical search), while the estimate
resulting from the polynomial regime resulted in unreliable and even
erroneous results.

We also obtained a specific class of Hamiltonian interpolations for the
search problem. To this end, we employed a recently developed theory, based on
the geometry of adiabatic evolutions, for obtaining suitable adiabatic
interpolations. This theory separates the adiabatic error into geometric and
non-geometric parts, and minimizes the former. We discussed three special
cases of the resulting class of interpolations in detail: (i) linear,
(ii) constant-norm, and (iii) constant-gap interpolations.

Finding strategies for minimizing the required total time as a function of a
given resource (system size, for example) is a desirable goal for many
applications, and is also of fundamental importance for the control of
quantum systems. We demonstrated explicitly how by imposing fairly general
controllability assumptions, which should be experimentally
straightforward to realize in certain scenarios, one can achieve a
significant reduction of the adiabatic error, and hence improve algorithmic
performance. The method we used relied on a polynomial expansion of the
adiabatic error, and resulted in the suppression of polynomial terms in $1/T$
by requiring smoothness for interpolations at the initial and final times.
It is evident that controlling the interpolation in this manner, at only two
points, has a substantial advantage over instantaneous control of the
Hamiltonian along the entire evolution. However, we demonstrated that there
is an extra price to pay for this error reduction: the exponential regime
(if exists) is extended, but with a slower rate of decay. This, in turn, may
result in an overestimation of the run time of the algorithm for some values
of the error threshold. We proposed a measure for quantifying the
performance of an adiabatic interpolation with various controllability
properties. In some cases, we also suggested a remedy for the above problem.
This fix necessitated further control over the Hamiltonian interpolation,
directly related to the amount of accessible energy in the system.
The interplay between the degree of required
control over Hamiltonian interpolations and the run time needed for
achieving a given accuracy was thus clearly exhibited.

Although we focused on the quantum search problem, our methods and most of
our results are applicable (perhaps with minor modifications) to a wider
class of problems---as discussed in the text. Since a principal goal in
adiabatic quantum algorithms, adiabatic quantum transport, quantum
annealing, and other applications of the adiabatic theorem, is the design of
algorithms with favorable performance-resource tradeoff, we hope that our
results will be of use in related physical applications.

\textit{Acknowledgments.}---Supported by the National Science Foundation
under grants No. CCF-0726439, No. CCF-956-400, No. PHY-802678 and
No. PHY-803304 (to DAL).

\bibliographystyle{apsrev}

\begin{thebibliography}{68}
\expandafter\ifx\csname natexlab\endcsname\relax\def\natexlab#1{#1}\fi
\expandafter\ifx\csname bibnamefont\endcsname\relax
  \def\bibnamefont#1{#1}\fi
\expandafter\ifx\csname bibfnamefont\endcsname\relax
  \def\bibfnamefont#1{#1}\fi
\expandafter\ifx\csname citenamefont\endcsname\relax
  \def\citenamefont#1{#1}\fi
\expandafter\ifx\csname url\endcsname\relax
  \def\url#1{\texttt{#1}}\fi
\expandafter\ifx\csname urlprefix\endcsname\relax\def\urlprefix{URL }\fi
\providecommand{\bibinfo}[2]{#2}
\providecommand{\eprint}[2][]{\url{#2}}

\bibitem[{\citenamefont{Brooke et~al.}(1999)\citenamefont{Brooke, Bitko,
  Rosenbaum, and Aeppli}}]{Brooke:99}
\bibinfo{author}{\bibfnamefont{J.}~\bibnamefont{Brooke}},
  \bibinfo{author}{\bibfnamefont{D.}~\bibnamefont{Bitko}},
  \bibinfo{author}{\bibfnamefont{T.~F.} \bibnamefont{Rosenbaum}},
  \bibnamefont{and} \bibinfo{author}{\bibfnamefont{G.}~\bibnamefont{Aeppli}},
  \bibinfo{journal}{Science} \textbf{\bibinfo{volume}{284}},
  \bibinfo{pages}{779} (\bibinfo{year}{1999}).

\bibitem[{\citenamefont{Farhi et~al.}()\citenamefont{Farhi, Goldstone, Gutmann,
  and Sipser}}]{Farhi1}
\bibinfo{author}{\bibfnamefont{E.}~\bibnamefont{Farhi}},
  \bibinfo{author}{\bibfnamefont{J.}~\bibnamefont{Goldstone}},
  \bibinfo{author}{\bibfnamefont{S.}~\bibnamefont{Gutmann}}, \bibnamefont{and}
  \bibinfo{author}{\bibfnamefont{M.}~\bibnamefont{Sipser}},
  \bibinfo{note}{eprint arXiv:quant-ph/0001106}.

\bibitem[{\citenamefont{Farhi et~al.}(2001)\citenamefont{Farhi, Goldstone,
  Gutmann, Lapan, Lundgren, and Preda}}]{Farhi2}
\bibinfo{author}{\bibfnamefont{E.}~\bibnamefont{Farhi}},
  \bibinfo{author}{\bibfnamefont{J.}~\bibnamefont{Goldstone}},
  \bibinfo{author}{\bibfnamefont{S.}~\bibnamefont{Gutmann}},
  \bibinfo{author}{\bibfnamefont{J.}~\bibnamefont{Lapan}},
  \bibinfo{author}{\bibfnamefont{A.}~\bibnamefont{Lundgren}}, \bibnamefont{and}
  \bibinfo{author}{\bibfnamefont{D.}~\bibnamefont{Preda}},
  \bibinfo{journal}{Science} \textbf{\bibinfo{volume}{292}},
  \bibinfo{pages}{472} (\bibinfo{year}{2001}).

\bibitem[{\citenamefont{Aharonov et~al.}(2007)\citenamefont{Aharonov, van Dam,
  Kempe, Landau, Lloyd, and Regev}}]{AharonovAdiabatic}
\bibinfo{author}{\bibfnamefont{D.}~\bibnamefont{Aharonov}},
  \bibinfo{author}{\bibfnamefont{W.}~\bibnamefont{van Dam}},
  \bibinfo{author}{\bibfnamefont{J.}~\bibnamefont{Kempe}},
  \bibinfo{author}{\bibfnamefont{Z.}~\bibnamefont{Landau}},
  \bibinfo{author}{\bibfnamefont{S.}~\bibnamefont{Lloyd}}, \bibnamefont{and}
  \bibinfo{author}{\bibfnamefont{O.}~\bibnamefont{Regev}},
  \bibinfo{journal}{SIAM J. Comput.} \textbf{\bibinfo{volume}{37}},
  \bibinfo{pages}{166} (\bibinfo{year}{2007}).

\bibitem[{\citenamefont{Oliveira and Terhal}(2005)}]{Oliveira:05}
\bibinfo{author}{\bibfnamefont{R.}~\bibnamefont{Oliveira}} \bibnamefont{and}
  \bibinfo{author}{\bibfnamefont{B.}~\bibnamefont{Terhal}},
  \bibinfo{journal}{Quantum Inf. Comput.} \textbf{\bibinfo{volume}{8}},
  \bibinfo{pages}{0900} (\bibinfo{year}{2005}).

\bibitem[{\citenamefont{Mizel et~al.}(2007)\citenamefont{Mizel, Lidar, and
  Mitchell}}]{MLM}
\bibinfo{author}{\bibfnamefont{A.}~\bibnamefont{Mizel}},
  \bibinfo{author}{\bibfnamefont{D.~A.} \bibnamefont{Lidar}}, \bibnamefont{and}
  \bibinfo{author}{\bibfnamefont{M.}~\bibnamefont{Mitchell}},
  \bibinfo{journal}{Phys. Rev. Lett.} \textbf{\bibinfo{volume}{99}},
  \bibinfo{pages}{070502} (\bibinfo{year}{2007}).

\bibitem[{\citenamefont{Jordan et~al.}(2006)\citenamefont{Jordan, Farhi, and
  Shor}}]{Jordan:05}
\bibinfo{author}{\bibfnamefont{S.~P.} \bibnamefont{Jordan}},
  \bibinfo{author}{\bibfnamefont{E.}~\bibnamefont{Farhi}}, \bibnamefont{and}
  \bibinfo{author}{\bibfnamefont{P.~W.} \bibnamefont{Shor}},
  \bibinfo{journal}{Phys. Rev. A} \textbf{\bibinfo{volume}{74}},
  \bibinfo{pages}{052322} (\bibinfo{year}{2006}).

\bibitem[{\citenamefont{Lidar}(2008)}]{Lidar:AQC-DD}
\bibinfo{author}{\bibfnamefont{D.~A.} \bibnamefont{Lidar}},
  \bibinfo{journal}{Phys. Rev. Lett.} \textbf{\bibinfo{volume}{100}},
  \bibinfo{pages}{160506} (\bibinfo{year}{2008}).

\bibitem[{\citenamefont{Childs et~al.}(2001)\citenamefont{Childs, Farhi, and
  Preskill}}]{Childs:01}
\bibinfo{author}{\bibfnamefont{A.}~\bibnamefont{Childs}},
  \bibinfo{author}{\bibfnamefont{E.}~\bibnamefont{Farhi}}, \bibnamefont{and}
  \bibinfo{author}{\bibfnamefont{J.}~\bibnamefont{Preskill}},
  \bibinfo{journal}{Phys. Rev. A} \textbf{\bibinfo{volume}{65}},
  \bibinfo{pages}{012322} (\bibinfo{year}{2001}).

\bibitem[{\citenamefont{Sarandy and
  Lidar}(2005{\natexlab{a}})}]{SarandyLidar:05}
\bibinfo{author}{\bibfnamefont{M.~S.} \bibnamefont{Sarandy}} \bibnamefont{and}
  \bibinfo{author}{\bibfnamefont{D.~A.} \bibnamefont{Lidar}},
  \bibinfo{journal}{Phys. Rev. Lett.} \textbf{\bibinfo{volume}{95}},
  \bibinfo{pages}{250503} (\bibinfo{year}{2005}{\natexlab{a}}).

\bibitem[{\citenamefont{Amin et~al.}(2008)\citenamefont{Amin, Love, and
  Truncik}}]{AminThermalAQC}
\bibinfo{author}{\bibfnamefont{M.~H.~S.} \bibnamefont{Amin}},
  \bibinfo{author}{\bibfnamefont{P.~J.} \bibnamefont{Love}}, \bibnamefont{and}
  \bibinfo{author}{\bibfnamefont{C.~J.~S.} \bibnamefont{Truncik}},
  \bibinfo{journal}{Phys. Rev. Lett.} \textbf{\bibinfo{volume}{100}},
  \bibinfo{pages}{060503} (\bibinfo{year}{2008}).

\bibitem[{\citenamefont{Sarandy and
  Lidar}(2005{\natexlab{b}})}]{SarandyLidar:04}
\bibinfo{author}{\bibfnamefont{M.~S.} \bibnamefont{Sarandy}} \bibnamefont{and}
  \bibinfo{author}{\bibfnamefont{D.~A.} \bibnamefont{Lidar}},
  \bibinfo{journal}{Phys. Rev. A} \textbf{\bibinfo{volume}{71}},
  \bibinfo{pages}{012331} (\bibinfo{year}{2005}{\natexlab{b}}).

\bibitem[{\citenamefont{Oreshkov and Calsamiglia}(2010)}]{Oreshkov:10}
\bibinfo{author}{\bibfnamefont{O.}~\bibnamefont{Oreshkov}} \bibnamefont{and}
  \bibinfo{author}{\bibfnamefont{J.}~\bibnamefont{Calsamiglia}},
  \bibinfo{journal}{Phys. Rev. Lett.} \textbf{\bibinfo{volume}{105}},
  \bibinfo{pages}{050503} (\bibinfo{year}{2010}).

\bibitem[{\citenamefont{Harris et~al.}()\citenamefont{Harris, Johnson, Lanting,
  Berkley, Johansson, Bunyk, Tolkacheva, Ladizinsky, Ladizinsky, Oh
  et~al.}}]{2010arXiv1004.1628H}
\bibinfo{author}{\bibfnamefont{R.}~\bibnamefont{Harris}},
  \bibinfo{author}{\bibfnamefont{M.~W.} \bibnamefont{Johnson}},
  \bibinfo{author}{\bibfnamefont{T.}~\bibnamefont{Lanting}},
  \bibinfo{author}{\bibfnamefont{A.~J.} \bibnamefont{Berkley}},
  \bibinfo{author}{\bibfnamefont{J.}~\bibnamefont{Johansson}},
  \bibinfo{author}{\bibfnamefont{P.}~\bibnamefont{Bunyk}},
  \bibinfo{author}{\bibfnamefont{E.}~\bibnamefont{Tolkacheva}},
  \bibinfo{author}{\bibfnamefont{E.}~\bibnamefont{Ladizinsky}},
  \bibinfo{author}{\bibfnamefont{N.}~\bibnamefont{Ladizinsky}},
  \bibinfo{author}{\bibfnamefont{T.}~\bibnamefont{Oh}}, \bibnamefont{et~al.},
  \bibinfo{note}{eprint arXiv:1004.1628}.

\bibitem[{\citenamefont{Born and Fock}(1928)}]{Fock}
\bibinfo{author}{\bibfnamefont{M.}~\bibnamefont{Born}} \bibnamefont{and}
  \bibinfo{author}{\bibfnamefont{V.~A.} \bibnamefont{Fock}},
  \bibinfo{journal}{Z. Physik} \textbf{\bibinfo{volume}{51}},
  \bibinfo{pages}{165} (\bibinfo{year}{1928}).

\bibitem[{\citenamefont{Kato}(1950)}]{Kato}
\bibinfo{author}{\bibfnamefont{T.}~\bibnamefont{Kato}}, \bibinfo{journal}{J.
  Phys. Soc. Japan} \textbf{\bibinfo{volume}{5}}, \bibinfo{pages}{435}
  (\bibinfo{year}{1950}).

\bibitem[{\citenamefont{Messiah}(1999)}]{Messiah:book}
\bibinfo{author}{\bibfnamefont{A.}~\bibnamefont{Messiah}},
  \emph{\bibinfo{title}{Quantum Mechanics}} (\bibinfo{publisher}{Dover
  Publications}, \bibinfo{address}{New York}, \bibinfo{year}{1999}).

\bibitem[{\citenamefont{Teufel}(2003)}]{Teufel:book}
\bibinfo{author}{\bibfnamefont{S.}~\bibnamefont{Teufel}},
  \emph{\bibinfo{title}{Adiabatic Perturbation Theory in Quantum Dynamics}}
  (\bibinfo{publisher}{Springer-Verlag}, \bibinfo{address}{Berlin},
  \bibinfo{year}{2003}).

\bibitem[{\citenamefont{Marzlin and Sanders}(2004)}]{MarzlinSanders}
\bibinfo{author}{\bibfnamefont{K.-P.} \bibnamefont{Marzlin}} \bibnamefont{and}
  \bibinfo{author}{\bibfnamefont{B.~C.} \bibnamefont{Sanders}},
  \bibinfo{journal}{Phys. Rev. Lett.} \textbf{\bibinfo{volume}{93}},
  \bibinfo{pages}{160408} (\bibinfo{year}{2004}).

\bibitem[{\citenamefont{Tong et~al.}(2005)\citenamefont{Tong, Singh, Kwek, and
  Oh}}]{Tong:05}
\bibinfo{author}{\bibfnamefont{D.~M.} \bibnamefont{Tong}},
  \bibinfo{author}{\bibfnamefont{K.}~\bibnamefont{Singh}},
  \bibinfo{author}{\bibfnamefont{L.~C.} \bibnamefont{Kwek}}, \bibnamefont{and}
  \bibinfo{author}{\bibfnamefont{C.~H.} \bibnamefont{Oh}},
  \bibinfo{journal}{Phys. Rev. Lett.} \textbf{\bibinfo{volume}{95}},
  \bibinfo{pages}{110407} (\bibinfo{year}{2005}).

\bibitem[{\citenamefont{Du et~al.}(2008)\citenamefont{Du, Hu, Wang, Wu, Zhao,
  and Suter}}]{du:060403}
\bibinfo{author}{\bibfnamefont{J.}~\bibnamefont{Du}},
  \bibinfo{author}{\bibfnamefont{L.}~\bibnamefont{Hu}},
  \bibinfo{author}{\bibfnamefont{Y.}~\bibnamefont{Wang}},
  \bibinfo{author}{\bibfnamefont{J.}~\bibnamefont{Wu}},
  \bibinfo{author}{\bibfnamefont{M.}~\bibnamefont{Zhao}}, \bibnamefont{and}
  \bibinfo{author}{\bibfnamefont{D.}~\bibnamefont{Suter}},
  \bibinfo{journal}{Phys. Rev. Lett.} \textbf{\bibinfo{volume}{101}},
  \bibinfo{eid}{060403} (pages~\bibinfo{numpages}{4}) (\bibinfo{year}{2008}).

\bibitem[{\citenamefont{Sarandy et~al.}(2004)\citenamefont{Sarandy, Wu, and
  Lidar}}]{SarandyWuLidar:04}
\bibinfo{author}{\bibfnamefont{M.~S.} \bibnamefont{Sarandy}},
  \bibinfo{author}{\bibfnamefont{L.-A.} \bibnamefont{Wu}}, \bibnamefont{and}
  \bibinfo{author}{\bibfnamefont{D.~A.} \bibnamefont{Lidar}},
  \bibinfo{journal}{Quantum Inf. Proc.} \textbf{\bibinfo{volume}{3}},
  \bibinfo{pages}{331} (\bibinfo{year}{2004}).

\bibitem[{\citenamefont{Amin}(2009)}]{Amin}
\bibinfo{author}{\bibfnamefont{M.~H.~S.} \bibnamefont{Amin}},
  \bibinfo{journal}{Phys. Rev. Lett.} \textbf{\bibinfo{volume}{102}},
  \bibinfo{pages}{220401} (\bibinfo{year}{2009}).

\bibitem[{\citenamefont{Avron et~al.}(1987)\citenamefont{Avron, Seiler, and
  Yaffe}}]{Avron:871}
\bibinfo{author}{\bibfnamefont{J.~E.} \bibnamefont{Avron}},
  \bibinfo{author}{\bibfnamefont{R.}~\bibnamefont{Seiler}}, \bibnamefont{and}
  \bibinfo{author}{\bibfnamefont{L.~G.} \bibnamefont{Yaffe}},
  \bibinfo{journal}{Commun. Math. Phys.} \textbf{\bibinfo{volume}{110}},
  \bibinfo{pages}{33} (\bibinfo{year}{1987}).

\bibitem[{\citenamefont{Avron et~al.}(1993)\citenamefont{Avron, Seiler, and
  Yaffe}}]{Avron:872}
\bibinfo{author}{\bibfnamefont{J.~E.} \bibnamefont{Avron}},
  \bibinfo{author}{\bibfnamefont{R.}~\bibnamefont{Seiler}}, \bibnamefont{and}
  \bibinfo{author}{\bibfnamefont{L.~G.} \bibnamefont{Yaffe}},
  \bibinfo{journal}{Commun. Math. Phys.} \textbf{\bibinfo{volume}{156}},
  \bibinfo{pages}{679} (\bibinfo{year}{1993}).

\bibitem[{\citenamefont{Nenciu}(1993)}]{Nenciu:93}
\bibinfo{author}{\bibfnamefont{G.}~\bibnamefont{Nenciu}},
  \bibinfo{journal}{Commun. Math. Phys.} \textbf{\bibinfo{volume}{152}},
  \bibinfo{pages}{479} (\bibinfo{year}{1993}).

\bibitem[{\citenamefont{Martinez}(1994)}]{Martinez}
\bibinfo{author}{\bibfnamefont{A.}~\bibnamefont{Martinez}},
  \bibinfo{journal}{J. Math. Phys.} \textbf{\bibinfo{volume}{35}},
  \bibinfo{pages}{3889} (\bibinfo{year}{1994}).

\bibitem[{\citenamefont{Hagedorn and Joye}(2002)}]{HagedornJoye:02}
\bibinfo{author}{\bibfnamefont{G.~A.} \bibnamefont{Hagedorn}} \bibnamefont{and}
  \bibinfo{author}{\bibfnamefont{A.}~\bibnamefont{Joye}}, \bibinfo{journal}{J.
  Math. Anal. Appl.} \textbf{\bibinfo{volume}{267}}, \bibinfo{pages}{235}
  (\bibinfo{year}{2002}).

\bibitem[{\citenamefont{Jansen et~al.}(2007)\citenamefont{Jansen, Ruskai, and
  Seiler}}]{JRS}
\bibinfo{author}{\bibfnamefont{S.}~\bibnamefont{Jansen}},
  \bibinfo{author}{\bibfnamefont{M.-B.} \bibnamefont{Ruskai}},
  \bibnamefont{and} \bibinfo{author}{\bibfnamefont{R.}~\bibnamefont{Seiler}},
  \bibinfo{journal}{J. Math. Phys.} \textbf{\bibinfo{volume}{48}},
  \bibinfo{pages}{102111} (\bibinfo{year}{2007}).

\bibitem[{\citenamefont{O'Hara and O'Leary}(2008)}]{o'hara:042319}
\bibinfo{author}{\bibfnamefont{M.~J.} \bibnamefont{O'Hara}} \bibnamefont{and}
  \bibinfo{author}{\bibfnamefont{D.~P.} \bibnamefont{O'Leary}},
  \bibinfo{journal}{Phys. Rev. A} \textbf{\bibinfo{volume}{77}},
  \bibinfo{pages}{042319} (\bibinfo{year}{2008}).

\bibitem[{\citenamefont{Boixo and Somma}(2010)}]{Boixo}
\bibinfo{author}{\bibfnamefont{S.}~\bibnamefont{Boixo}} \bibnamefont{and}
  \bibinfo{author}{\bibfnamefont{R.~D.} \bibnamefont{Somma}},
  \bibinfo{journal}{Phys. Rev. A} \textbf{\bibinfo{volume}{81}},
  \bibinfo{pages}{032308} (\bibinfo{year}{2010}).

\bibitem[{\citenamefont{Lidar et~al.}(2009)\citenamefont{Lidar, Rezakhani, and
  Hamma}}]{LRH}
\bibinfo{author}{\bibfnamefont{D.~A.} \bibnamefont{Lidar}},
  \bibinfo{author}{\bibfnamefont{A.~T.} \bibnamefont{Rezakhani}},
  \bibnamefont{and} \bibinfo{author}{\bibfnamefont{A.}~\bibnamefont{Hamma}},
  \bibinfo{journal}{J. Math. Phys.} \textbf{\bibinfo{volume}{50}},
  \bibinfo{pages}{102106} (\bibinfo{year}{2009}).

\bibitem[{\citenamefont{Grover}(1997)}]{Grover}
\bibinfo{author}{\bibfnamefont{L.~K.} \bibnamefont{Grover}},
  \bibinfo{journal}{Phys. Rev. Lett.} \textbf{\bibinfo{volume}{79}},
  \bibinfo{pages}{325} (\bibinfo{year}{1997}).

\bibitem[{\citenamefont{Rezakhani et~al.}(2009)\citenamefont{Rezakhani, Kuo,
  Hamma, Lidar, and Zanardi}}]{QAB}
\bibinfo{author}{\bibfnamefont{A.~T.} \bibnamefont{Rezakhani}},
  \bibinfo{author}{\bibfnamefont{W.-J.} \bibnamefont{Kuo}},
  \bibinfo{author}{\bibfnamefont{A.}~\bibnamefont{Hamma}},
  \bibinfo{author}{\bibfnamefont{D.~A.} \bibnamefont{Lidar}}, \bibnamefont{and}
  \bibinfo{author}{\bibfnamefont{P.}~\bibnamefont{Zanardi}},
  \bibinfo{journal}{Phys. Rev. Lett.} \textbf{\bibinfo{volume}{103}},
  \bibinfo{pages}{080502} (\bibinfo{year}{2009}).

\bibitem[{\citenamefont{Nielsen and Chuang}(2000)}]{Nielsen:book}
\bibinfo{author}{\bibfnamefont{M.~A.} \bibnamefont{Nielsen}} \bibnamefont{and}
  \bibinfo{author}{\bibfnamefont{I.~L.} \bibnamefont{Chuang}},
  \emph{\bibinfo{title}{Quantum Computation and Quantum Information}}
  (\bibinfo{publisher}{Cambridge University Press},
  \bibinfo{address}{Cambridge, England}, \bibinfo{year}{2000}).

\bibitem[{\citenamefont{Roland and Cerf}(2002)}]{RolandCerf}
\bibinfo{author}{\bibfnamefont{J.}~\bibnamefont{Roland}} \bibnamefont{and}
  \bibinfo{author}{\bibfnamefont{N.~J.} \bibnamefont{Cerf}},
  \bibinfo{journal}{Phys. Rev. A} \textbf{\bibinfo{volume}{65}},
  \bibinfo{pages}{042308} (\bibinfo{year}{2002}).

\bibitem[{\citenamefont{Andrecut and Ali}(2007)}]{Andrecut}
\bibinfo{author}{\bibfnamefont{M.}~\bibnamefont{Andrecut}} \bibnamefont{and}
  \bibinfo{author}{\bibfnamefont{M.~K.} \bibnamefont{Ali}},
  \bibinfo{journal}{Intl. J. Theor. Phys.} \textbf{\bibinfo{volume}{2}},
  \bibinfo{pages}{447} (\bibinfo{year}{2007}).

\bibitem[{\citenamefont{Zalka}(1999)}]{Zalka:97}
\bibinfo{author}{\bibfnamefont{C.}~\bibnamefont{Zalka}},
  \bibinfo{journal}{Phys. Rev. A} \textbf{\bibinfo{volume}{60}},
  \bibinfo{pages}{2746} (\bibinfo{year}{1999}).

\bibitem[{\citenamefont{Grover}(1998)}]{Grover-2}
\bibinfo{author}{\bibfnamefont{L.~K.} \bibnamefont{Grover}},
  \bibinfo{journal}{Phys. Rev. Lett.} \textbf{\bibinfo{volume}{80}},
  \bibinfo{pages}{4329} (\bibinfo{year}{1998}).

\bibitem[{\citenamefont{Boyer et~al.}(1998)\citenamefont{Boyer, Brassard,
  Hoyer, and Tapp}}]{Boyer:96}
\bibinfo{author}{\bibfnamefont{M.}~\bibnamefont{Boyer}},
  \bibinfo{author}{\bibfnamefont{G.}~\bibnamefont{Brassard}},
  \bibinfo{author}{\bibfnamefont{P.}~\bibnamefont{Hoyer}}, \bibnamefont{and}
  \bibinfo{author}{\bibfnamefont{A.}~\bibnamefont{Tapp}},
  \bibinfo{journal}{Fortschr. Phys.} \textbf{\bibinfo{volume}{46}},
  \bibinfo{pages}{493} (\bibinfo{year}{1998}).

\bibitem[{\citenamefont{Grover}(2005)}]{Grover-3}
\bibinfo{author}{\bibfnamefont{L.~K.} \bibnamefont{Grover}},
  \bibinfo{journal}{Phys. Rev. Lett.} \textbf{\bibinfo{volume}{95}},
  \bibinfo{pages}{150501} (\bibinfo{year}{2005}).

\bibitem[{\citenamefont{Biham et~al.}(1999)\citenamefont{Biham, Biham, Biron,
  Grassl, and Lidar}}]{G-1}
\bibinfo{author}{\bibfnamefont{E.}~\bibnamefont{Biham}},
  \bibinfo{author}{\bibfnamefont{O.}~\bibnamefont{Biham}},
  \bibinfo{author}{\bibfnamefont{D.}~\bibnamefont{Biron}},
  \bibinfo{author}{\bibfnamefont{M.}~\bibnamefont{Grassl}}, \bibnamefont{and}
  \bibinfo{author}{\bibfnamefont{D.~A.} \bibnamefont{Lidar}},
  \bibinfo{journal}{Phys. Rev. A} \textbf{\bibinfo{volume}{60}},
  \bibinfo{pages}{2742} (\bibinfo{year}{1999}).

\bibitem[{\citenamefont{Biham et~al.}(2000)\citenamefont{Biham, Biham, Biron,
  Grassl, Lidar, and Shapira}}]{Lidar:PRA01Grover}
\bibinfo{author}{\bibfnamefont{E.}~\bibnamefont{Biham}},
  \bibinfo{author}{\bibfnamefont{O.}~\bibnamefont{Biham}},
  \bibinfo{author}{\bibfnamefont{D.}~\bibnamefont{Biron}},
  \bibinfo{author}{\bibfnamefont{M.}~\bibnamefont{Grassl}},
  \bibinfo{author}{\bibfnamefont{D.~A.} \bibnamefont{Lidar}}, \bibnamefont{and}
  \bibinfo{author}{\bibfnamefont{D.}~\bibnamefont{Shapira}},
  \bibinfo{journal}{Phys. Rev. A} \textbf{\bibinfo{volume}{63}},
  \bibinfo{pages}{012310} (\bibinfo{year}{2000}).

\bibitem[{\citenamefont{Accardi and Sabbadini}()}]{G-2}
\bibinfo{author}{\bibfnamefont{L.}~\bibnamefont{Accardi}} \bibnamefont{and}
  \bibinfo{author}{\bibfnamefont{R.}~\bibnamefont{Sabbadini}},
  \bibinfo{note}{eprint arXiv:quant-ph/0012143.}

\bibitem[{\citenamefont{Chuang et~al.}(1998)\citenamefont{Chuang, Gershenfeld,
  and Kubinec}}]{Chuang}
\bibinfo{author}{\bibfnamefont{I.~L.} \bibnamefont{Chuang}},
  \bibinfo{author}{\bibfnamefont{N.}~\bibnamefont{Gershenfeld}},
  \bibnamefont{and} \bibinfo{author}{\bibfnamefont{M.}~\bibnamefont{Kubinec}},
  \bibinfo{journal}{Phys. Rev. Lett.} \textbf{\bibinfo{volume}{80}},
  \bibinfo{pages}{3408} (\bibinfo{year}{1998}).

\bibitem[{\citenamefont{Jones et~al.}(1998)\citenamefont{Jones, Mosca, and
  Hansen}}]{Jones}
\bibinfo{author}{\bibfnamefont{J.~A.} \bibnamefont{Jones}},
  \bibinfo{author}{\bibfnamefont{M.}~\bibnamefont{Mosca}}, \bibnamefont{and}
  \bibinfo{author}{\bibfnamefont{R.~H.} \bibnamefont{Hansen}},
  \bibinfo{journal}{Nature} \textbf{\bibinfo{volume}{393}},
  \bibinfo{pages}{344} (\bibinfo{year}{1998}).

\bibitem[{\citenamefont{Ollerenshaw et~al.}(2003)\citenamefont{Ollerenshaw,
  Lidar, and Kay}}]{Ollerenshaw:02}
\bibinfo{author}{\bibfnamefont{J.}~\bibnamefont{Ollerenshaw}},
  \bibinfo{author}{\bibfnamefont{D.~A.} \bibnamefont{Lidar}}, \bibnamefont{and}
  \bibinfo{author}{\bibfnamefont{L.~E.} \bibnamefont{Kay}},
  \bibinfo{journal}{Phys. Rev. Lett.} \textbf{\bibinfo{volume}{91}},
  \bibinfo{pages}{217904} (\bibinfo{year}{2003}).

\bibitem[{\citenamefont{Daems and Gu\'{e}rin}(2007)}]{Daems}
\bibinfo{author}{\bibfnamefont{D.}~\bibnamefont{Daems}} \bibnamefont{and}
  \bibinfo{author}{\bibfnamefont{S.}~\bibnamefont{Gu\'{e}rin}},
  \bibinfo{journal}{Phys. Rev. Lett.} \textbf{\bibinfo{volume}{99}},
  \bibinfo{pages}{170503} (\bibinfo{year}{2007}).

\bibitem[{\citenamefont{Ivanov et~al.}(2010)\citenamefont{Ivanov, Ivanov,
  Linington, and Vitanov}}]{Ivanov}
\bibinfo{author}{\bibfnamefont{S.~S.} \bibnamefont{Ivanov}},
  \bibinfo{author}{\bibfnamefont{P.~A.} \bibnamefont{Ivanov}},
  \bibinfo{author}{\bibfnamefont{I.~E.} \bibnamefont{Linington}},
  \bibnamefont{and} \bibinfo{author}{\bibfnamefont{N.~V.}
  \bibnamefont{Vitanov}}, \bibinfo{journal}{Phys. Rev. A}
  \textbf{\bibinfo{volume}{81}}, \bibinfo{pages}{042328}
  (\bibinfo{year}{2010}).

\bibitem[{\citenamefont{Rezakhani et~al.}(2010)\citenamefont{Rezakhani, Abasto,
  Lidar, and Zanardi}}]{AQC-intrinsic}
\bibinfo{author}{\bibfnamefont{A.~T.} \bibnamefont{Rezakhani}},
  \bibinfo{author}{\bibfnamefont{D.~F.} \bibnamefont{Abasto}},
  \bibinfo{author}{\bibfnamefont{D.~A.} \bibnamefont{Lidar}}, \bibnamefont{and}
  \bibinfo{author}{\bibfnamefont{P.}~\bibnamefont{Zanardi}},
  \bibinfo{journal}{Phys. Rev. A} \textbf{\bibinfo{volume}{82}},
  \bibinfo{pages}{012321} (\bibinfo{year}{2010}).

\bibitem[{\citenamefont{Aharonov and Ta-Shma}(2007)}]{AharonovTa-Shma}
\bibinfo{author}{\bibfnamefont{D.}~\bibnamefont{Aharonov}} \bibnamefont{and}
  \bibinfo{author}{\bibfnamefont{A.}~\bibnamefont{Ta-Shma}},
  \bibinfo{journal}{SIAM J. Comput.} \textbf{\bibinfo{volume}{37}},
  \bibinfo{pages}{47} (\bibinfo{year}{2007}).

\bibitem[{\citenamefont{Siu}(2007)}]{Siu}
\bibinfo{author}{\bibfnamefont{M.~S.} \bibnamefont{Siu}},
  \bibinfo{journal}{Phys. Rev. A} \textbf{\bibinfo{volume}{75}},
  \bibinfo{pages}{062337} (\bibinfo{year}{2007}).

\bibitem[{\citenamefont{Garrido and Sancho}(1962)}]{Sancho}
\bibinfo{author}{\bibfnamefont{L.~M.} \bibnamefont{Garrido}} \bibnamefont{and}
  \bibinfo{author}{\bibfnamefont{F.~J.} \bibnamefont{Sancho}},
  \bibinfo{journal}{Physica A} \textbf{\bibinfo{volume}{28}},
  \bibinfo{pages}{553} (\bibinfo{year}{1962}).

\bibitem[{\citenamefont{Latorre and Or\'{u}s}(2004)}]{Latorre}
\bibinfo{author}{\bibfnamefont{J.~I.} \bibnamefont{Latorre}} \bibnamefont{and}
  \bibinfo{author}{\bibfnamefont{R.}~\bibnamefont{Or\'{u}s}},
  \bibinfo{journal}{Phys. Rev. A} \textbf{\bibinfo{volume}{69}},
  \bibinfo{pages}{062302} (\bibinfo{year}{2004}).

\bibitem[{\citenamefont{Schaller et~al.}(2006)\citenamefont{Schaller, Mostame,
  and Sch\"{u}tzhold}}]{Schaller}
\bibinfo{author}{\bibfnamefont{G.}~\bibnamefont{Schaller}},
  \bibinfo{author}{\bibfnamefont{S.}~\bibnamefont{Mostame}}, \bibnamefont{and}
  \bibinfo{author}{\bibfnamefont{R.}~\bibnamefont{Sch\"{u}tzhold}},
  \bibinfo{journal}{Phys. Rev. A} \textbf{\bibinfo{volume}{73}},
  \bibinfo{pages}{062307} (\bibinfo{year}{2006}).

\bibitem[{\citenamefont{Sachdev}(1999)}]{Sachdev:book}
\bibinfo{author}{\bibfnamefont{S.}~\bibnamefont{Sachdev}},
  \emph{\bibinfo{title}{Quantum Phase Transitions}}
  (\bibinfo{publisher}{Cambridge University Press},
  \bibinfo{address}{Cambridge, England}, \bibinfo{year}{1999}).

\bibitem[{\citenamefont{Sch\"{u}tzhold and Schaller}(2006)}]{Schutzhold:06}
\bibinfo{author}{\bibfnamefont{R.}~\bibnamefont{Sch\"{u}tzhold}}
  \bibnamefont{and} \bibinfo{author}{\bibfnamefont{G.}~\bibnamefont{Schaller}},
  \bibinfo{journal}{Phys. Rev. A} \textbf{\bibinfo{volume}{74}},
  \bibinfo{pages}{060304(R)} (\bibinfo{year}{2006}).

\bibitem[{\citenamefont{Schaller}(2008)}]{Schaller2}
\bibinfo{author}{\bibfnamefont{G.}~\bibnamefont{Schaller}},
  \bibinfo{journal}{Phys. Rev. A} \textbf{\bibinfo{volume}{78}},
  \bibinfo{pages}{032328} (\bibinfo{year}{2008}).

\bibitem[{\citenamefont{Amin and Choi}(2009)}]{AminChoi}
\bibinfo{author}{\bibfnamefont{M.~H.~S.} \bibnamefont{Amin}} \bibnamefont{and}
  \bibinfo{author}{\bibfnamefont{V.}~\bibnamefont{Choi}},
  \bibinfo{journal}{Phys. Rev. A} \textbf{\bibinfo{volume}{80}},
  \bibinfo{pages}{062326} (\bibinfo{year}{2009}).

\bibitem[{\citenamefont{Galindo and Pascual}(1990)}]{Galindo1:book}
\bibinfo{author}{\bibfnamefont{A.}~\bibnamefont{Galindo}} \bibnamefont{and}
  \bibinfo{author}{\bibfnamefont{P.}~\bibnamefont{Pascual}},
  \emph{\bibinfo{title}{Quantum Mechanics I}}
  (\bibinfo{publisher}{Springer-Verlag}, \bibinfo{address}{Berlin},
  \bibinfo{year}{1990}).

\bibitem[{\citenamefont{Pfeifer}(1993)}]{ETU-1}
\bibinfo{author}{\bibfnamefont{P.}~\bibnamefont{Pfeifer}},
  \bibinfo{journal}{Phys. Rev. Lett.} \textbf{\bibinfo{volume}{70}},
  \bibinfo{pages}{3365} (\bibinfo{year}{1993}).

\bibitem[{\citenamefont{Giovannetti et~al.}(2003)\citenamefont{Giovannetti,
  Lloyd, and Maccone}}]{ETU-2}
\bibinfo{author}{\bibfnamefont{V.}~\bibnamefont{Giovannetti}},
  \bibinfo{author}{\bibfnamefont{S.}~\bibnamefont{Lloyd}}, \bibnamefont{and}
  \bibinfo{author}{\bibfnamefont{L.}~\bibnamefont{Maccone}},
  \bibinfo{journal}{Phys. Rev. A} \textbf{\bibinfo{volume}{67}},
  \bibinfo{pages}{052109} (\bibinfo{year}{2003}).

\bibitem[{\citenamefont{Levitin and Toffoli}(2009)}]{ETU-3}
\bibinfo{author}{\bibfnamefont{L.~B.} \bibnamefont{Levitin}} \bibnamefont{and}
  \bibinfo{author}{\bibfnamefont{T.}~\bibnamefont{Toffoli}},
  \bibinfo{journal}{Phys. Rev. Lett.} \textbf{\bibinfo{volume}{103}},
  \bibinfo{pages}{160502} (\bibinfo{year}{2009}).

\bibitem[{\citenamefont{Avron et~al.}()\citenamefont{Avron, Fraas, Graf, and
  Grech}}]{LZ-Avron}
\bibinfo{author}{\bibfnamefont{J.~E.} \bibnamefont{Avron}},
  \bibinfo{author}{\bibfnamefont{M.}~\bibnamefont{Fraas}},
  \bibinfo{author}{\bibfnamefont{G.~M.} \bibnamefont{Graf}}, \bibnamefont{and}
  \bibinfo{author}{\bibfnamefont{P.}~\bibnamefont{Grech}},
  \bibinfo{note}{eprint arXiv:0912.4640.}

\bibitem[{\citenamefont{Arfken and Weber}(2001)}]{Arfken:book}
\bibinfo{author}{\bibfnamefont{G.~B.} \bibnamefont{Arfken}} \bibnamefont{and}
  \bibinfo{author}{\bibfnamefont{H.~J.} \bibnamefont{Weber}},
  \emph{\bibinfo{title}{Mathematical Methods for Physicists}}
  (\bibinfo{publisher}{Academic Press}, \bibinfo{address}{San Diego},
  \bibinfo{year}{2001}).

\bibitem[{\citenamefont{Hayek}(2001)}]{Hayek:book}
\bibinfo{author}{\bibfnamefont{S.~I.} \bibnamefont{Hayek}},
  \emph{\bibinfo{title}{Advanced Mathematical Methods in Science and
  Engineering}} (\bibinfo{publisher}{Marcel Dekker}, \bibinfo{address}{New
  York}, \bibinfo{year}{2001}).

\bibitem[{\citenamefont{Nakahara}(2003)}]{Nakahara:book}
\bibinfo{author}{\bibfnamefont{M.}~\bibnamefont{Nakahara}},
  \emph{\bibinfo{title}{Geometry, Topology and Physics}}
  (\bibinfo{publisher}{Institute of Physics}, \bibinfo{address}{Bristol and
  Philadelphia}, \bibinfo{year}{2003}).

\bibitem[{\citenamefont{Gel'fand and Shilov}(1964)}]{Gelfand:book}
\bibinfo{author}{\bibfnamefont{I.~M.} \bibnamefont{Gel'fand}} \bibnamefont{and}
  \bibinfo{author}{\bibfnamefont{G.~E.} \bibnamefont{Shilov}},
  \emph{\bibinfo{title}{Generalized Functions, Vol. 1: Properties and
  Operations}} (\bibinfo{publisher}{Academic Press}, \bibinfo{address}{New
  York}, \bibinfo{year}{1964}).

\end{thebibliography}


\end{document}